\documentclass[sigconf,dvipsnames]{acmart}

\usepackage{amsmath}
\usepackage[disabled]{aplcomments}
\usepackage{comment}
\usepackage{xspace}
\usepackage{hyperref}
\usepackage{url}
\usepackage{listings}
\usepackage{graphicx}
\usepackage{caption}
\usepackage[labelformat=simple]{subcaption}
\usepackage{multirow}
\usepackage{adjustbox}
\usepackage{array}
\usepackage{xcolor}
\usepackage{booktabs}
\usepackage{tikz}
\usepackage{pbox}
\usepackage{wrapfig}
\usepackage{paralist}
\usepackage{pdfpages}
\usepackage{makecell}

\copyrightyear{2018} 
\acmYear{2018} 
\setcopyright{acmcopyright}
\acmConference[SIGMOD'18]{2018 International Conference on Management of Data}{June 10--15, 2018}{Houston, TX, USA}
\acmBooktitle{SIGMOD'18: 2018 International Conference on Management of Data, June 10--15, 2018, Houston, TX, USA}
\acmPrice{15.00}
\acmDOI{10.1145/3183713.3196891}
\acmISBN{978-1-4503-4703-7/18/06}

\newcommand{\tr}[1]{\llbracket #1 \rrbracket}

\newcommand{\figlabel}[1]{\label{f:#1}}
\newcommand{\seclabel}[1]{\label{s:#1}}
\newcommand{\tablabel}[1]{\label{t:#1}}

\newcommand{\secref}[1]{\S\ref{s:#1}}
\newcommand{\figref}[1]{Figure~\ref{f:#1}}
\newcommand{\tabref}[1]{Table~\ref{t:#1}}
\renewcommand{\eqref}[1]{Equation~\ref{eq:#1}}
\newcommand\sys{\textsc{Casper}\xspace}
\newcommand\biglam{Big$\lambda$\xspace}

\newcommand\suites{7\xspace}
\newcommand\totalcf{101\xspace}
\newcommand\totalbench{55\xspace}
\newcommand\transcf{82\xspace}

\newcommand*{\pscomment}[1]{\textcolor{OliveGreen}{// #1}}
\newcolumntype{P}[1]{>{\centering\arraybackslash}m{#1}}
\newcolumntype{Q}[1]{>{\arraybackslash}m{#1}}

\definecolor{javapurple}{rgb}{0.5,0,0.35}
\definecolor{javagreen}{rgb}{0,0.4,0}

\lstdefinelanguage{java}
{
  morekeywords={for, int, static, class, public, void, if, new, return, RDD, else, then, double, List, choose, generator},
  morecomment=[l]{//},
  morecomment=[s]{/*}{*/},
  morestring=[b]",
  basicstyle=\small\ttfamily,
  numbers=left, firstnumber=1, numberstyle=\tiny\color{gray},
  showstringspaces=false,
  escapeinside={(*@}{@*)},
  keywordstyle=\color{javapurple},
  columns=fullflexible,
  showlines=true,
  xleftmargin=0.5cm,
  moredelim=**[is][\color{NavyBlue}]{!}{!},
  moredelim=**[is][\color{OliveGreen}]{<_}{_>}
}

\lstdefinelanguage{dafny}
{
  morekeywords={function, seq, int, method, reads, requires, in, forall, if, then, else, ensures},
  morecomment=[l]{//},
  morecomment=[s]{/*}{*/},
  morestring=[b]",
  basicstyle=\footnotesize\ttfamily,
  escapeinside={@}{@},
  numbers=left, firstnumber=1, numberstyle=\tiny\color{gray},
  showstringspaces=false,
  keywordstyle=\color{javapurple},
  columns=fullflexible,
  xleftmargin=0.6cm  
}

\lstdefinelanguage{psuedo}
{
  morekeywords={continue, elif, for, if, else, forall, function, while, in, not, break, is, do, then, return, and},
  morestring=[b]",
  basicstyle=\footnotesize\ttfamily,
  escapeinside={@}{@},
  escapechar=\&,
  showstringspaces=false,
  keywordstyle=\color{javapurple},
  columns=fullflexible,
  numbers=left, firstnumber=1, numberstyle=\footnotesize\color{gray},
  tabsize=1,
  xleftmargin=0.6cm,
  moredelim=**[is][\color{BrickRed}]{@}{@},
  moredelim=**[is][\color{BlueViolet}]{!}{!}
}

\newcommenter{alvin}{0.4,1.0,1.0}
\newcommenter{maaz}{1.0,0.4,1.0}

\begin{document}

\setlength{\pdfpageheight}{\paperheight}
\setlength{\pdfpagewidth}{\paperwidth}

\title{Automatically Leveraging MapReduce Frameworks for Data-Intensive Applications}

\author{Maaz Bin Safeer Ahmad}
\affiliation{
  %\department{Computer Science \& Engineering}
  \institution{University of Washington \\
  {\tt maazsaf@cs.washington.edu} }
%  \city{Seattle}
%  \state{Washington}
%  \country{USA}
}
%\email{maazsaf@cs.washington.edu}

\author{Alvin Cheung}
\affiliation{
  %\department{Computer Science \& Engineering}
  \institution{University of Washington \\
  {\tt akcheung@cs.washington.edu} }
%  \city{Seattle}
%  \state{Washington}
%  \country{USA}
}
%\email{akcheung@cs.washington.edu}

\subtitle{{\tt casper.uwplse.org}}

\begin{abstract}
MapReduce is a popular programming paradigm for developing large-scale, data-intensive computation. Many frameworks that implement this paradigm have recently been developed. To leverage these frameworks, however, developers must become familiar with their APIs and rewrite existing code. We present \sys, a new tool that automatically translates sequential Java programs into the MapReduce paradigm. \sys identifies potential code fragments to rewrite and translates them in two steps: (1) \sys uses {\it program synthesis} to search for a program summary (i.e., a functional specification) of each code fragment. The summary is expressed using a high-level intermediate language resembling the MapReduce paradigm and verified to be semantically equivalent to the original using a theorem prover. (2) \sys generates executable code from the summary, using either the Hadoop, Spark, or Flink API. We evaluated \sys by automatically converting real-world, sequential Java benchmarks to MapReduce. The resulting benchmarks perform up to 48.2$\times$ faster compared to the original.
\end{abstract}

\maketitle

\section{Introduction}
MapReduce~\cite{mapreduce}, a popular paradigm for developing data-intensive applications, has varied and highly efficient implementations~\cite{hadoop, spark, flink, disco}. All these implementations expose an application programming interface (API) to developers. While the concrete syntax differs slightly across the different APIs, they all require developers to organize their computation into {\em map} and {\em reduce} stages in order to leverage their optimizations.

While exposing optimization via an API shields application developers from the complexities of distributed computing, this approach contains a major drawback: for legacy applications to leverage MapReduce frameworks, developers must first understand the existing code's function and subsequently re-organize the computation using mappers and reducers. Similarly, novice programmers, unfamiliar with the MapReduce paradigm, must first learn the different APIs in order to express their computation accordingly. Both require a significant expenditure of time and effort. Further, each code rewrite or algorithm reformulation opens another opportunity to introduce bugs.

One way to alleviate these issues is to build a compiler that translates code written in another paradigm (e.g., imperative code) into MapReduce. Classical compilers, like logical to physical query plan compilers~\cite{hyper}, use pattern matching rules, i.e., the compilers contain a number of rules that recognize different input code patterns (e.g., a sequential loop over lists) and translate the matched code fragment into the target (e.g., a single-stage map and reduce). As in query compilers, designing the rules is challenging: they must be both {\em correct}, i.e., the translated code should have the same semantics as the input, and sufficiently {\em expressive} to capture the wide variety of coding patterns that developers use to express their computations. We are aware of only one such compiler that translates imperative Java programs into MapReduce~\cite{mold}, and the number of rules involved in that compiler makes it difficult to maintain and modify.

This paper describes a new tool, \sys, that translates sequential Java code into semantically equivalent MapReduce programs. Rather than relying on rules to translate different code patterns, \sys is inspired by prior work on {\em cost-based query optimization}~\cite{selingerOptimizer}, which considers compilation to be a dynamic search problem. However, given that the inputs are general-purpose programs, the space of possible target programs is much larger than it is for query optimization. To address this issue, \sys leverages recent advances in program synthesis~\cite{sumitDimensionsInSynthesis, RasSynthesisSurvey} to search for MapReduce programs into which it can rewrite a given input sequential Java code fragment. To reduce the search space, \sys searches over the space of {\it program summaries}, which are expressed using a {\em high-level intermediate language (IR)} that we designed. As we discuss in~\secref{sec_3.1_speclang}, the IR's design succinctly expresses computations in the MapReduce paradigm yet remains sufficiently easy to translate into the concrete syntax of the target API.

To search for summaries, \sys first performs lightweight program analysis to generate a description of the space of MapReduce programs that a given input code fragment {\it might} be equivalent to. The search space is also described using our high-level IR. \sys then uses an off-the-shelf {\it program synthesizer} to perform the search, but it is guided by an {\em incremental search algorithm} and our {\em domain-specific cost model} to speed the process. A {\em theorem prover} is used to check whether the found program summary is indeed semantically equivalent to the input. Once proved, the summary is translated into the concrete syntax of the target MapReduce API. Since the performance of the translated program often depends on input data characteristics (e.g., skewness), \sys generates multiple semantically equivalent MapReduce programs for a given input and produces a monitor module that switches among them based on runtime statistics; the monitor and switcher are automatically generated during compilation.

Compared to prior approaches, \sys does not require compiler developers to design or maintain any pattern matching rules. Furthermore, the entire translation process is completely automatic. We evaluated \sys using a number of benchmarks and real-world Java applications and demonstrated both \sys's ability to translate an input program into MapReduce equivalents and the significant performance improvements that result.

In summary, our paper makes the following contributions:

\begin{asparaitem}
\item We propose a new high-level intermediate representation (IR) to express the semantics of sequential Java programs in the MapReduce paradigm. The language is succinct to be easily translated into multiple MapReduce APIs, yet expressive to describe the semantics of many real-world benchmarks written in a general-purpose language. Furthermore, programs written in our IR can be automatically checked for correctness using a theorem prover (\secref{sec_4.1_verifierfailures}). The IR, being a high-level language, also lets us perform various {\it semantic optimizations} using our cost model (\secref{sec_5_optimal}).

\item We describe an efficient search technique for program summaries expressed in the IR without requiring any pattern matching rules. Our technique is both {\em sound} and {\em complete} with respect to the input search space. Unlike classical compilers, which rely on pattern matching to drive translation, our technique leverages program synthesis to dynamically search for summaries. Our technique is novel in that it incrementally searches for summaries based on cost. It also uses verification failures to systematically prune the search space and a hierarchy of search grammars to speed the summary search. This lets us translate benchmarks that have not been translated in any prior work (\secref{sec_4.1_verifierfailures}).

\item There are often multiple ways to express the same input as MapReduce programs. Therefore, our technique can generate multiple semantically equivalent MapReduce versions of the input. It also automatically inserts code that collects statistics during program execution to adaptively switch among the different generated versions (\secref{sec_5_runtimemodule}).

\item We implemented our methodology in \sys, a tool that converts sequential Java programs into three MapReduce implementations: Spark, Hadoop, and Flink. We evaluated the feasibility and effectiveness of \sys by translating real-world benchmarks from \suites different suites from multiple domains. Across \totalbench benchmarks, \sys translated \transcf of \totalcf code fragments. The translated benchmarks performed up to 48.2$\times$ faster compared to the original ones and were competitive even with other distributed implementations, including manual ones (\secref{evaluation}).
\end{asparaitem}

\section{Overview}
\seclabel{sec_2_overview}

This section describes how we model the MapReduce programming paradigm and demonstrates by example how \sys translates sequential code into MapReduce programs. 

%%%%%%%%%%%%%%%%%%%%%%%%%%%%%%%%%%%%%%% 2.1 %%%%%%%%%%%%%%%%%%%%%%%%%%%%%%%%%%%%%%%

\subsection{MapReduce Operators}
\seclabel{sec_2.1_mapreduce}

MapReduce organizes computation using two operators: {\em map} and {\em reduce}. The map operator has the following type signature:
\begin{align*}
\boldsymbol{map} &: (mset[\tau] , \: \lambda_m)  \; \longrightarrow \; mset[(\kappa,\nu)]  \\
\boldsymbol{\lambda_m} &: \tau \; \longrightarrow \; mset[(\kappa,\nu)]
\end{align*}
Input into $map$ is a multiset (i.e., bag) of type $\tau$ and a unary transformer function $\lambda_m$, which converts a value of type $\tau$ into a multiset of key-value pairs of types $\kappa$ and $\nu$. The map operator then concurrently applies $\lambda_m$ to every element in the multiset and returns the union of all multisets generated by $\lambda_m$.
\begin{align*}
\boldsymbol{reduce} &: (mset[(\kappa,\nu)], \: \lambda_r) \longrightarrow mset[(\kappa,\nu)] \\
\boldsymbol{\lambda_r} &: (\nu, \nu) \longrightarrow \nu
\end{align*}
Input into $reduce$ is a multiset of key-value pairs and a binary transformer function $\lambda_r$, which combines two values of type $\nu$ to produce a final value. The reduce operator first groups all key-value pairs by key (also known as shuffling) and then uses $\lambda_r$ to combine, in parallel, the bag of values for each key-group into a single value. The output of $reduce$ is another multiset of key-value pairs, where each pair holds a unique key. If the transformer function $\lambda_r$ is commutative and associative, then $reduce$ can be further optimized by concurrently applying $\lambda_r$ to pairs of values in a key-group.

\sys's goal is to translate a sequential code fragment into a MapReduce program that is expressed using the $map$ and $reduce$ operators. The challenges in doing so are: (1) identify the correct sequence of operators to apply, and (2) implement the corresponding transformer functions. We next discuss how \sys overcomes these challenges.

%%%%%%%%%%%%%%%%%%%%%%%%%%%%%%%%%%%%%%% 2.2 %%%%%%%%%%%%%%%%%%%%%%%%%%%%%%%%%%%%%%%

\subsection{Translating Imperative Code to MapReduce}
\seclabel{sec_2.2_example}

\begin{figure}[t]
\centering
\begin{subfigure}[t]{\linewidth}
\begin{lstlisting}[language=java]
!@Summary( (*@ \label{lst:pcStart} @*)
  $\color{NavyBlue}m = map(reduce(map(mat,\lambda_{m1}),\lambda_{r}),\lambda_{m2})$
  $\color{NavyBlue}\lambda_{m1} : (i, j, v) \rightarrow \{(i, v)\}$
  $\color{NavyBlue}\lambda_{r} : (v_1, v_2) \rightarrow v_1 + v_2$
  $\color{NavyBlue}\lambda_{m2} : (k, v) \rightarrow \{(k, v/cols)\}$ )! (*@ \label{lst:pcEnd} @*)
int[] rwm(int[][] mat, int rows, int cols) {
    int[] m = new int[rows];
    for (int i = 0; i < rows; i++) { (*@ \label{lst:outerLoopStart} @*)
        int sum = 0;
        for (int j = 0; j < cols; j++) (*@ \label{lst:innerLoopStart} @*)
            sum += mat[i][j]; (*@ \label{lst:innerLoopEnd} @*)
        m[i] = sum / cols; 
    } (*@ \label{lst:outerLoopEnd} @*)
    return m; 
}
\end{lstlisting}
\vspace{-0.5cm}
\caption{Input: Sequential Java code}
\vspace{0.1cm}
\figlabel{rwm_java}
\end{subfigure}
\begin{subfigure}[t]{\linewidth}
\begin{lstlisting}[language=java]
RDD rwm(RDD mat, int rows, int cols) {
  RDD m = mat.mapToPair(e -> new Tuple(e.i, e.v));
  m = m.reduceByKey((v1, v2) -> (v1 + v2));
  m = m.mapValues(v -> (v / cols));
  return m; 
}
\end{lstlisting}
\vspace{-0.5cm}
\caption{Output: Apache Spark code}
%\vspace{0.2cm}
\figlabel{rwm_spark}
\end{subfigure}
\vspace{-0.4cm}
\caption{Using \sys to translate the row-wise mean benchmark to MapReduce (Spark).}
\vspace{-0.4cm}
\end{figure}

\sys takes in Java code with loop nests that sequentially iterate over data and translates the code into a semantically equivalent MapReduce program to be executed by the target framework. To demonstrate, we show how \sys translates a real-word benchmark from the Phoenix suite~\cite{pheonix}.

As shown in~\figref{rwm_java}, the benchmark takes as input a matrix ({\tt mat}) and computes, using nested loops, the column vector ({\tt m}) containing the mean value of each row in the matrix. Assume the code is annotated with a {\em program summary} that helps with the translation into MapReduce. The program summary, written using a high-level intermediate representation (IR), describes how the output of the code fragment (i.e., {\tt m}) can be computed using a series of {\em map} and {\em reduce} stages from the input data (i.e., {\tt mat}), as shown in lines~\ref{lst:pcStart} to \ref{lst:pcEnd} in \figref{rwm_java}. While the summary is not executable, translating from that into the concrete syntax of a MapReduce framework (say, Spark) would be much easier than translating from the original input code. This is shown in~\figref{rwm_spark} where the {\em map} and {\em reduce} primitives from our summary are translated into the corresponding Spark API calls.

Unfortunately, the input code does not have such a summary, which must therefore be inferred. \sys does this via program synthesis and verification, as we explain in~\secref{sec_3_synthesis}.

%%%%%%%%%%%%%%%%%%%%%%%%%%%%%%%%%%%%%%% 2.3 %%%%%%%%%%%%%%%%%%%%%%%%%%%%%%%%%%%%%%%

\subsection{System Architecture}
\seclabel{sec_2.3_architecture}

\begin{figure}
    \centering
    \includegraphics[width=1\linewidth]{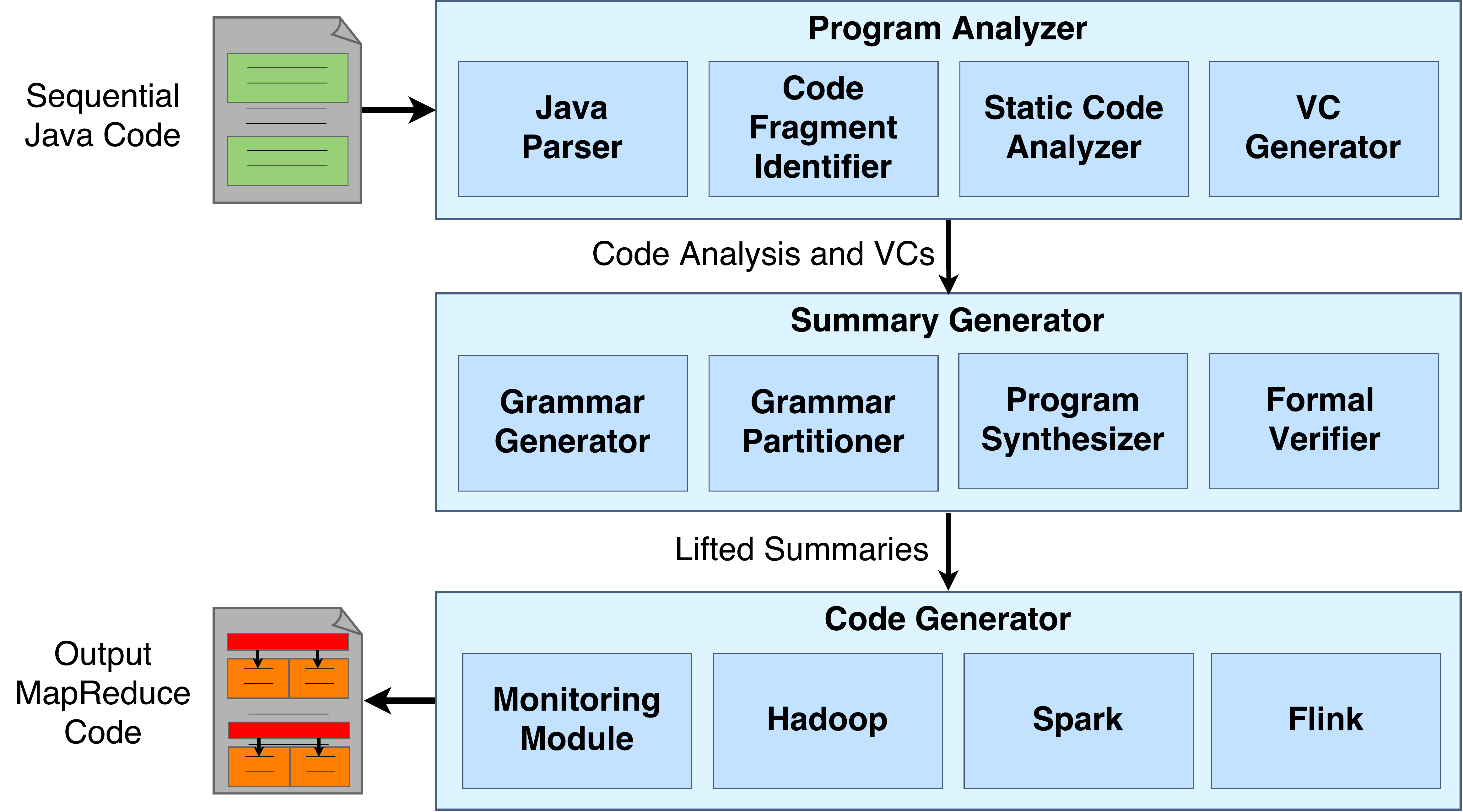}
    \vspace{-0.7cm}
    \caption{\sys's system architecture. Sequential code fragments 
    (green) are translated into MapReduce tasks (orange).}
    \vspace{-0.3cm}
    \figlabel{arch}
\end{figure}

\figref{arch} shows \sys's overall design. We now discuss the three primary modules that make up \sys's compilation pipeline.

First, the {\em program analyzer} parses the input code into an Abstract Syntax Tree (AST) and uses static program analysis to identify code fragments for translation (\secref{sec_6.1_limitations}). In addition, for each identified code fragment, it prepares: (1) a {\it search space description} encoded using our high-level IR that lets the synthesizer search for a valid program summary (\secref{sec_3.1_speclang}), and (2) {\it verification conditions} (VCs) (\secref{sec_3.2_verifcond}) to automatically ascertain that the induced program summary is semantically equivalent to the input.

Next, the {\em summary generator} synthesizes and verifies program summaries (\secref{sec_3.3_searchstrat} and \secref{sec_4.1_verifierfailures}). To speed up the search, it partitions the search space so that it can be efficiently traversed using our incremental synthesis algorithm (\secref{sec_4.2_incremental}).

Once a summary is inferred, the {\em code generator} translates it into executable code. \sys currently supports three MapReduce frameworks: Spark, Hadoop, and Flink. Additionally, this component also generates code that collects data statistics to adaptively choose among different implementations during runtime (\secref{sec_5_runtimemodule}).

\section{Synthesizing Program Summaries}
\seclabel{sec_3_synthesis}

As discussed, \sys discovers a program summary for each code fragment before translation. Technically, a program summary is a {\em postcondition}~\cite{hoare} of the input code that describes the program state after the code fragment is executed. 
% Much research has been done on inferring postconditions, and \sys uses program synthesis to for this task. 
In this section, we explain: (1) the IR \sys uses to express summaries, (2) how \sys verifies a summary's validity, and (3) the search algorithm \sys uses to find valid summaries given a search space description.
%from the space of all possible postconditions expressible in the IR.

%%%%%%%%%%%%%%%%%%%%%%%%%%%%%%%%%%%%%%% 3.1 %%%%%%%%%%%%%%%%%%%%%%%%%%%%%%%%%%%%%%%
% The objective of this subsection is to introduce our spec language. Key ideas to
% convey are: 1) pcs must describe output variables, 2) map reduce are known but 
% transformers are not. 3) dynamic grammar generation, 4) recursive bounds
% 5) hint towards incremental grammar gen without explaining anything.

\subsection{A High-level IR for Program Summaries} 
\seclabel{sec_3.1_speclang}

One approach to synthesize summaries directly searches in programs written in the target framework's API. This does not scale well; Spark alone offers over 80 high-level operators, even though many of them have similar semantics and differ only in their implementation or syntax (e.g., map, flatMap, filter). To speed up synthesis, we instead search in programs written in a high-level IR that abstracts away syntactical differences and describes only the functionality of a few essential operators. The goals of the IR are: (1) to express summaries that are translatable into the target API, and (2) to let the synthesizer efficiently search for summaries that are equivalent to the input program. To address these goals, \sys's IR models two MapReduce primitives that are similar to the {\em map} and {\em fold} operators in Haskell (see \secref{sec_2.1_mapreduce}). In addition, our IR models the {\em join} primitive, which takes as input two multisets of key-value pairs and returns all pairs of elements with matching keys. The IR does not currently model the full range of operators across different MapReduce implementations; however, it already lets \sys capture a wide array of computations expressible using the paradigm and is sufficiently general to be translatable into different MapReduce APIs while keeping the search problem tractable, as we demonstrate in~\secref{evaluation}.

\begin{figure}[t]
\begin{small}
\begin{subfigure}[t]{1\linewidth}
\footnotesize
\begin{eqnarray*}
PS &:=& \forall v. \;\; v = MR \; \mid \; \forall v. \;\; v = MR[v_{id}] \\
MR &:=& map(MR,\lambda_m) \; \mid \;  reduce(MR,\lambda_r) 
	\; \mid \; join(MR,MR) \; \mid \; data \\ 
\lambda_m &:=& f : (val) \rightarrow \{ Emit \} \\
\lambda_r &:=& f : (val_1, val_2) \rightarrow Expr \\
Emit &:=& emit(Expr,Expr) \; \mid \; if(Expr) \; emit(Expr,Expr) \; \mid \\
	 && if(Expr) \; emit(Expr,Expr) \;\; else \; Emit \\
Expr &:=& Expr \;\; op \;\; Expr \; \mid \; op \;\; Expr \;\; \mid \; f(Expr,Expr,...) \; \mid \\
	 && n \; \mid \; var \; \mid \; (Expr, Expr)
\end{eqnarray*}
\vspace{-0.2cm}
\end{subfigure}
\end{small}
\vspace{-0.4cm}
\noindent\rule{\linewidth}{0.4pt}
\begin{subfigure}[t]{1\linewidth}
\begin{footnotesize}
\begin{subfigure}[t]{0.5\linewidth}
\begin{eqnarray*}
v &\in& Output~Variables \\ 
op &\in& Operators
\end{eqnarray*}
\end{subfigure}
\begin{subfigure}[t]{0.5\linewidth}
\begin{eqnarray*}
v_{id} &\in& Variable~ID, \\ 
f &\in& Library~Methods
\end{eqnarray*}
\end{subfigure}
\end{footnotesize}
\end{subfigure}
\vspace{-0.3cm}
\caption{Excerpt of the IR for program summaries (PSs), a full description of which is provided in Appendix~\ref{appendix:ir}.}
\vspace{-0.3cm}
\figlabel{sec_3.1_speclang}
\end{figure}

\figref{sec_3.1_speclang} shows a subset of \sys's IR, used to express both program summaries and the search space. The IR assumes that program summaries are expressed in the stylized form shown in~\figref{sec_3.1_speclang} as {\em PS}, which states that each output variable $v$ (i.e., a variables updated in the code fragment), must be computed using a sequence of $map$, $reduce$ and $join$ operations over the inputs (e.g., the arrays or collections being iterated). While doing so ensures that the summary is translatable into the target API, the implementations of $\lambda_m$ and $\lambda_r$ for the $map$ and $reduce$ operators depend on the code fragment being translated. We leave these functions to be synthesized and restrict the body of $\lambda_m$ to a sequence of $emit$ statements, where each $emit$ statement produces a single key-value pair, and the body of $\lambda_r$ is an expression that evaluates to a single value of the required type. Besides $emit$, the bodies of $\lambda_m$ and $\lambda_r$'s can consist of conditionals and other operations on tuples, as shown in~\figref{sec_3.1_speclang}. The output of the MapReduce expression is an associative array of key-value pairs; the unique key $v_{id}$ for each variable is used to access the computed value of that variable. Appendix \ref{appendix:ir} lists the full set of types and operators that our IR supports.

%The synthesizer will decide on both the correct number of $Emit$ statements to use, what constitutes the key and value for each emitted key-value pair. 

%Our IR also supports conditionals, allowing the $map$ stage to model filter operations. In addition to primitive data-types, the language supports tuples as well. The output of each $reduce$ expression is an associative array of key-value pairs, with the variable ID $v_{id}$ of each output variable as the key mapping to its computed value. The variable ID for an output variable is a unique integer assigned by \sys during compilation that allows extracting the computed values for each variable from the MapReduce program output.

\subsection{Defining the Search Space}

In addition to program summaries, \sys also uses the IR to describe the search space of summaries for the synthesizer. It does so by generating a {\em grammar} for each input code fragment, like the one shown in~\figref{sec_3.1_speclang}. The synthesizer traverses the grammar by expanding on each production rule and checks whether any generated candidate constitutes a valid summary (as explained in~\secref{sec_3.2_verifcond}).

To generate the search space grammar, \sys analyzes the input code to extract the following properties and their type information:

\begin{compactenum}
\item Variables in scope at the beginning of the input code
\item Variables that are modified within the input code
\item The operators and library methods used
\end{compactenum}

The code analyzer extracts these properties using standard program analyses. It computes (1) and (2) using live variable and dataflow analysis~\cite{dragonBook}, and it computes (3) by scanning functions that are invoked in the input code. We currently assume that input variables are not aliased to each other and put guards on the translated code to ensure that is the case.\footnote{\footnotesize Thus, if variable handles {\tt v1} and {\tt v2} are both inputs into the same code fragment, \sys wraps the translated code as: {\tt if (v1 != v2) \{ [\sys translated code] \} else \{ [original code] \}}. Computing precise alias information requires more engineering~\cite{alias} and does not impact our approach.}   Appendix~\ref{appendix:example} shows the analysis results for the TPC-H Q6 benchmark, and we discuss the limitations of our program analyzer module implementation in \secref{sec_6.1_limitations}.

Given this information, the summary generator builds a search space grammar that is specialized to the code fragment being translated. \figref{sec_4.2_eqClassFig} shows sample grammars that are generated for the code shown in~\figref{rwm_java}.\footnote{Refer to Appendix~\ref{appendix:example} to see how a grammar can be encoded in our IR.} The input code uses addition and division; hence, the grammar includes addition and division in its production rules for $\lambda_m$ and $\lambda_r$. Furthermore, \sys also uses type information of variables to prune invalid production rules in the grammar. For instance, if the output variable $v$ is of type $int$, the final operation in the synthesized MapReduce expression must evaluate to a value of type $int$. Since the output type of a reduce operation is inferred from the type of its input, we can propagate this information to restrict the type of values the reduce operation accepts. To make synthesis tractable and the search space finite, \sys imposes recursive bounds on the production rules. For instance, it limits the number of MapReduce operations a program summary can use and the number of $emit$ statements in a single transformer function. In \secref{sec_4.2_incremental}, we discuss how \sys further specializes the search space %for each input code fragment
by changing the set of production rules available in the grammar or specifying different recursive bounds.

%%%%%%%%%%%%%%%%%%%%%%%%%%%%%%%%%%%%%%% 3.1 %%%%%%%%%%%%%%%%%%%%%%%%%%%%%%%%%%%%%%%
% The objective of this subsection is to explain how Cora can verify a given
% pc. We explain what VCs are and then explain that VCs need invariants. Use the
% rwn as an example, finally formulate the search problem to connect with the next
% section.

\subsection{Verifying Program Summaries}
\seclabel{sec_3.2_verifcond}

To search for a valid summary within the search space, \sys requires a way to check whether a candidate summary is semantically equivalent to the input code. It does so using standard techniques in program verification, namely, by creating {\em verification conditions} based on Hoare logic~\cite{hoare}. Verification conditions are Boolean predicates that, given a program statement $S$ and a postcondition (i.e., program summary) $P$, state what must be true {\em before} $S$ is executed in order for $P$ to be a valid postcondition of $S$. Verification conditions can be systematically generated for imperative program statements, including those processed by \sys~\cite{semofpl,  vcgen}. However, each loop statement requires an extra {\em loop invariant} to construct an inductive proof. Loop invariants are Boolean predicates that are true before and after every execution of the loop body regardless of how many times the loop executes. 

The general problem of inferring the strongest loop invariants or postconditions is undecidable~\cite{vcgen,semofpl}. Unlike prior work, however, two factors make our problem solvable: first, our summaries are restricted to only those expressible using the IR described in~\secref{sec_3.1_speclang}, which lacks many problematic features (e.g., pointers) that a general-purpose language would have. Moreover, we are interested only in finding loop invariants that are {\em strong enough} to establish the validity of the synthesized program summaries.

\begin{figure}[t]
\centering
\begin{subfigure}[t]{\linewidth}
\begin{align*}
in&variant(m,i) \equiv \;\; 0 \leq i \leq rows \;\; \wedge \\ 
&m = map(reduce(map(mat[0..i],\lambda_{m1}),\lambda_{r}),\lambda_{m2})
\end{align*}
\vspace{-0.5cm}
\caption{Outer loop invariant}
\figlabel{rwm_inv}
\end{subfigure}
\begin{subfigure}[t]{\linewidth}
\vspace{0.2cm}
\centering
\begin{tabular}{ rl }
  Initiation & \pbox{20cm}{
      \vspace{0.1cm}
      $(i=0) \rightarrow Inv(m, i)$
      \vspace{0.1cm}
  } \\
  \hline
  Continuation & \pbox{20cm}{
      \vspace{0.1cm}
      $Inv(m,i) \wedge (i<rows) \rightarrow$ \\ $Inv(m[i \mapsto sum(mat[i])/cols], i+1)$
      \vspace{0.1cm}
  } \\
  \hline
  Termination & \pbox{20cm}{
      \vspace{0.1cm}
      $Inv(m,i) \wedge \neg(i<rows) \rightarrow PS(m, i)$
      \vspace{0.1cm}
  } \\
\end{tabular}
\vspace{-0.1cm}
\caption{Verification conditions to ascertain the correctness of the program summary $PS$ given loop invariant $Inv$}
\figlabel{rwm_vcs}
\end{subfigure}
\vspace{-0.3cm}
\caption{Proof of soundness for the row-wise mean benchmark.}
\vspace{-0.3cm}
\end{figure}

As an example, \figref{rwm_inv} shows an outer loop invariant $Inv$, which can be used to prove the validty of the program summary shown in~\figref{rwm_java}. \figref{rwm_vcs} shows the verification conditions \sys constructs to state what the program summary and invariant must satisfy. We can check that this loop invariant and program summary are indeed valid based on Hoare logic as follows. First, the {\em initiation} clause asserts that the invariant holds before the loop, i.e., when {\tt i} is zero. This is true because the invariant asserts that the MapReduce expression is true only for the first $i$ rows of the input matrix. Hence, when {\tt i} is zero, the MapReduce expression is executed on an empty dataset, and the output value for each row is 0. Next, the {\em continuation} clause asserts that after one more execution of the loop body, the $i^{th}$ index of output vector {\tt m} should hold the mean for the $i^{th}$ row of the matrix {\tt mat}. This is true since the value of {\tt i} is incremented inside the loop body, which implies that the mean for the $i^{th}$ row has been computed. Finally, the {\em termination} condition completes the proof by asserting that if the invariant is true, and {\tt i} has reached the end of the matrix, then the program summary $PS$ must now hold as well. This is true since {\tt i} now equals the number of rows in the matrix, and the loop invariant asserts that {\tt m} equals the MapReduce expression executed over the entire matrix, which is the same assertion as our program summary.

\sys formulates the search problem for finding program summaries by constructing the verification conditions for the given code fragment and leaving the body of the summary (and any necessary invariants for loops) to be synthesized. For the program summary and invariants, the search space is expressed using the same IR as discussed in~\secref{sec_3.1_speclang}. Formally, the synthesis problem is:
\begin{align}
\exists \;ps,\;inv_1,\ldots,inv_n. \;\; \forall \sigma. \; VC(P, ps, inv_1, \ldots, inv_n, \sigma)
\label{synthformulae}
\end{align}
In other words, \sys's goal is to find a program summary $ps$ and any required invariants $inv_1, \ldots, inv_n$ such that for all possible program states $\sigma$, the verification conditions for the input code fragment $P$ are true. After the synthesizer has identified a candidate summary and invariants, \sys sends them and the verification conditions to a theorem prover  (see~\secref{sec_4.1_verifierfailures}), and to the code generator to generate executable MapReduce code if the program summary is proven to be correct.

%%%%%%%%%%%%%%%%%%%%%%%%%%%%%%%%%%%%%%% 3.1 %%%%%%%%%%%%%%%%%%%%%%%%%%%%%%%%%%%%%%%
% Explain what SuGuS is. Explain 2-step verification. Explain algorithm using
% code walkthrough.

\subsection{Search Strategy}
\seclabel{sec_3.3_searchstrat}

\sys uses an off-the-shelf program synthesizer, Sketch~\cite{sketch}, to infer program summaries and loop invariants. Sketch takes as input: (1) a set of candidate summaries and invariants encoded as a grammar (e.g., \figref{sec_3.1_speclang}), and (2) the correctness specification for the summary in the form of verification conditions. It then attempts to find a program summary (and any invariants needed) using the provided grammar such that the verification conditions hold true.

The universal quantifier in Eq.\ref{synthformulae} make the synthesis problem  challenging. Therefore, \sys uses a two-step process to ensure that the found summary is valid. First, it leverages Sketch's {\em bounded model checking} to verify the candidate program summary over a finite (i.e., ``bounded'') subset of all possible program states. For example, \sys restricts the maximum size of the input dataset and the range of values for integer inputs. Finding a solution for this weakened specification can be done very efficiently by the synthesizer. Once a candidate program summary can be verified for the bounded domain, \sys passes the summary to a theorem prover to determine its soundness over the entire domain, which is more expensive computationally. \sys currently translates the summary along with an automatically generated proof script to Dafny~\cite{dafny} for full verification. This two-step verification makes \sys's synthesis algorithm sound, without compromising efficiency. 
%Bounded checking done by the synthesizer acts as a filter to quickly eliminate all obviously incorrect program summaries. The more computationally expensive verification offered by the theorem prover is reserved only for the set of program summaries generated by the synthesizer in the first step.

%\input{sec_3.3_cegisFig.tex}

\subsubsection{Synthesis Algorithm}
%Counter-Example Guided Inductive Synthesis}
\seclabel{sec_3.3.1_cegis}

\figref{sec_4.3_fullalgorithm} (lines \ref{lst:cegisStart} to \ref{lst:cegisEnd}) shows the core CEGIS~\cite{sketching} algorithm \sys's synthesizer uses. The algorithm is an iterative interaction between two modules: a candidate program summary generator and a bounded model checker. The candidate summary generator takes as input the IR grammar $G$, the verification conditions for the input code fragment $VC$, and a set of concrete program states $\Phi$. To start the process, the synthesizer populates $\Phi$ with a few randomly chosen states, and generates program summary candidate $ps$ and any needed invariants $inv_1,\ldots,inv_n$ from $G$ such that $\forall \sigma \in \Phi~.~ VC(ps, inv_1,\ldots,inv_n, \sigma)$ is true. Next, the bounded model checker verifies whether the candidate program summary holds over the bounded domain. If it does, the algorithm returns $ps$ as the solution. Otherwise, the model checker returns a counter-example state $\phi$ such that $VC(ps,inv_1,\ldots,inv_n,\phi)$ is false. The algorithm adds $\phi$ to $\Phi$ and restarts the program summary generator. This continues until either a program summary is found that passes bounded model checking or the search space is exhausted.

\begin{figure}
\begin{lstlisting}[language=psuedo]
function synthesize (G, VC): &\label{lst:cegisStart}&
  $\Phi$ = {} &\pscomment{set of random program states}&
  while true do
    ps, inv$_{1..n}$ = generateCandidate(G, VC, $\Phi$)   &\label{lst:genCandCall}& 
    if ps is null then return null &\pscomment{search space exhausted}& &\label{lst:spaceexhausted}&
    $\phi$ = boundedVerify(ps, inv$_{1..n}$, VC)  &\label{lst:bmcCall}&
    if $\phi$ is null then return (ps, inv$_{1..n}$) &\pscomment{summary found}&
    else $\Phi = \Phi \; \cup \phi$  &\pscomment{counter-example found}& &\label{lst:sumfound}& &\label{lst:cegisEnd}&

function findSummary ($A$, VC):
  G = generateGrammar($A$)
  $\Gamma$ = generateClasses(G) &\label{lst:genGrammar}&
  $\Omega$ = {} &\pscomment{summaries that failed verification}&
  $\Delta$ = {} &\pscomment{summaries that passed verification}&
  for $\gamma \in \Gamma$ do &\label{lst:genGrammarSt}&
    while true do 
      c = synthesize($\gamma$ - $\Omega$ - $\Delta$, VC) &\label{lst:callSynth}&
      if c is null and $\Delta$ is null then 
        break &\pscomment{move to next grammar class}&
      else if c is null then 
        return $\Delta$ &\pscomment{search complete}& &\label{lst:solsfound}&
      else if fullVerify(c, VC) then $\Delta = \Delta \; \cup$ c &\label{lst:fullVerify}&
      else $\Omega = \Omega \; \cup$ c
  return null &\pscomment{no solution found}& &\label{lst:nosols}&
\end{lstlisting}
\vspace{-0.4cm}
\caption{\sys's search algorithm.}
\vspace{-0.4cm}
\figlabel{sec_4.3_fullalgorithm}
\end{figure}

%, i.e., 
%
% \begin{align*}
% \forall ps, inv_1,\ldots,inv_n \in G. \; \exists \sigma \in \Phi. \; \neg VC(ps, inv_1,\ldots,\allowbreak inv_n, \sigma)    
% \end{align*}

A limitation of the CEGIS algorithm is that, while efficient, the found program summary might be true only for the finite domain and thus will be rejected by the theorem prover when checking for validity over the entire domain. In this case, \sys dynamically changes the search space grammar to exclude the candidate program summary that does not verify and restarts the synthesizer to generate a new candidate summary using the preceding algorithm. We discuss this process in detail in~\secref{sec_4.1_verifierfailures}.

\section{Improving Summary Search}
\seclabel{sec_4_improvingsynth}
We now discuss the techniques \sys uses to make the search for program summaries more robust and efficient.

%%%%%%%%%%%%%%%%%%%%%%%%%%%%%%%%%%%%%%% 4.1 %%%%%%%%%%%%%%%%%%%%%%%%%%%%%%%%%%%%%%%
% Explain how to handle verifier failrues i.e. completeness of search

\begin{figure*}[t]
\begin{minipage}[t]{0.32\linewidth}
\footnotesize
\begin{subfigure}[b]{1\linewidth}
{%\setlength\extrarowheight{10pt}
\begin{tabular}[b]{ |c|c|c|c| }
\hline
    \textbf{Property} &
	$G_1$ & 
	$G_2$ & 	
	$G_3$  \\	
\hline
\makecell{Map/Reduce \\ Sequence} & m & m $\rightarrow$ r & m $\rightarrow$ r $\rightarrow$ m \\
\hline
\makecell{\# Emits \\in $\lambda_m$}  & 1 & 2 & 2 \\
\hline
\makecell{Key/Value \\ Type} & int & \makecell{int} & \makecell{int or \\ Tuple<int,int>} \\ 
\hline
%\makecell{Max \\Cost} & 32N & 192N &  400N \\ 
%\hline
\end{tabular}
}
\vspace{0.4cm}
\end{subfigure}
\end{minipage}
\begin{minipage}[t]{0.67\linewidth}
\begin{subfigure}{1\linewidth}
\includegraphics[width=1\linewidth]{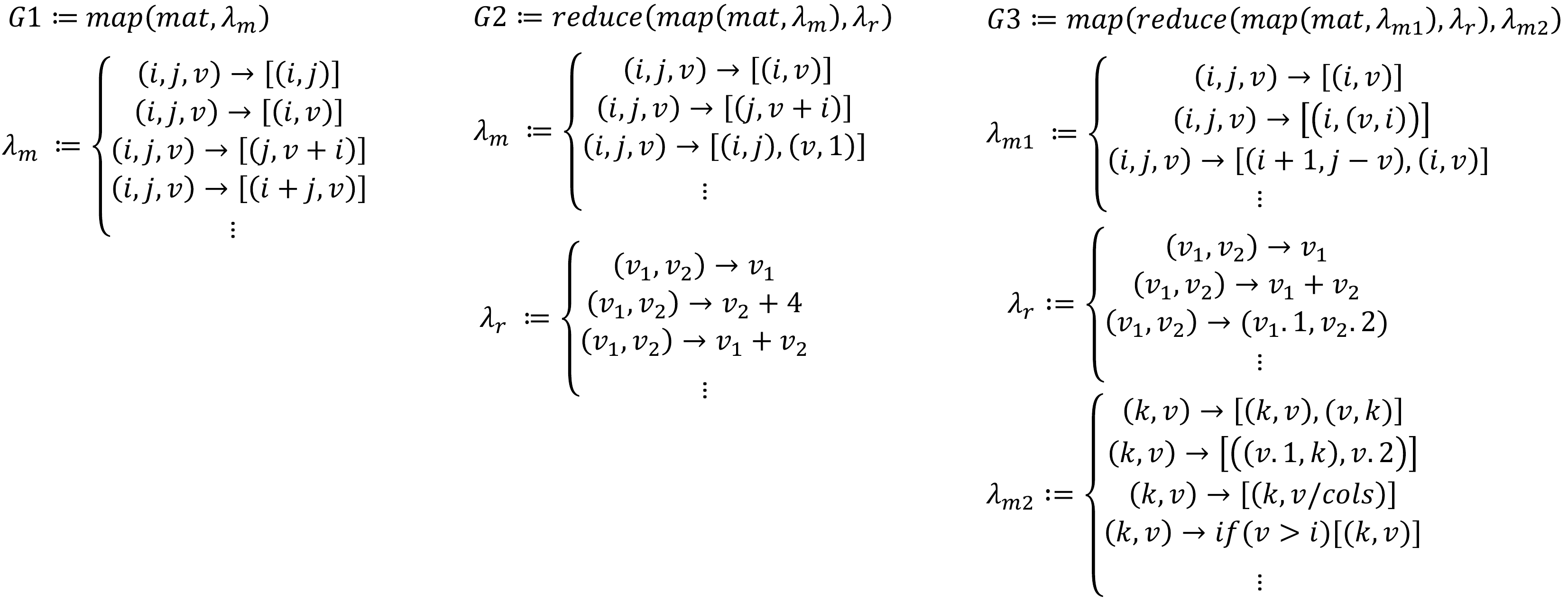}
\end{subfigure}
\end{minipage}
\vspace{-0.5cm}
\caption{Incremental grammar generation. \sys generates a hierarchy of grammars to optimize search.}
\vspace{-0.3cm}
\figlabel{sec_4.2_eqClassFig}
\end{figure*}

\subsection{Leveraging Verifier Failures}
\seclabel{sec_4.1_verifierfailures}

As mentioned, the program summary that the synthesizer returns can fail theorem prover validation due to the bounded domain used during search. For instance, assume we bound the integer inputs to have a maximum value of 4 in the synthesizer. In this bounded domain, the expressions {\tt v} and {\tt Math.min(4,v)} (where {\tt v} is an input integer) are deemed to be equivalent even though they are not equal in practice. While prior work~\cite{qbs, stng} simply fails to translate such benchmarks if the theorem prover rejects the candidate summary, \sys uses a two-phase verification technique to eliminate such candidates. This ensures that \sys's search is complete with respect to the search space defined by the grammar. 

To achieve completeness, \sys must first prevent  summaries that failed the theorem prover from being regenerated by the synthesizer. A naive approach would be to restart the synthesizer until a new summary is found, assuming that the algorithm implemented by the synthesizer is non-deterministic. However, this approach is incomplete because the algorithm may never terminate since it can continually return the same incorrect summary. Instead, \sys modifies the search space to ensure forward progress. Recall from~\secref{sec_3.3_searchstrat} that the search space for candidate summaries $\{c_1, \ldots, c_n\}$ is specified using an input grammar that is generated by the program analyzer and passed to the synthesizer. Thus, to prevent a candidate $c_f$ that fails the theorem prover from being repeatedly generated from grammar $G$, \sys simply passes in a new grammar $G - \{c_f\}$ to the synthesizer. This is implemented by passing additional constraints to the synthesizer to block a summary from being regenerated.
\\[0.3em]
\noindent\textbf{Theorem.} \sys's algorithm for inferring program summaries is sound and complete with respect to the given search space.
\\
A proof sketch for this theorem is provided in Appendix~\ref{appendix:proof}.

\figref{sec_4.3_fullalgorithm} shows how \sys infers program summaries and invariants. \sys calls the synthesizer to generate a candidate summary {\tt c} on line \ref{lst:callSynth} and attempts to verify {\tt c} by passing it to the theorem prover on line \ref{lst:fullVerify}. If verification fails, {\tt c} is added to $\Omega$, the set of incorrect summaries, and the synthesizer is restarted with a new grammar $G - \Omega$. We explain the full algorithm in \secref{sec_4.3_fullalgo}.

%In practice, this approach works well. 
In \secref{twophaseverifeval}, we provide experimental results that illustrate how our two-phase verification algorithm effectively finds program summaries even when faced with verification failures.

% highlight the necessity of having a two phase verification in making search for the correct summary efficient, and how doing so enables us to translate code fragments that prior work is unable to do due to verifier failures.

%%%%%%%%%%%%%%%%%%%%%%%%%%%%%%%%%%%%%%% 4.1 %%%%%%%%%%%%%%%%%%%%%%%%%%%%%%%%%%%%%%%
% Explain incremental grammar generation, motivation is to improve search speed

\subsection{Incremental Grammar Generation}
\seclabel{sec_4.2_incremental}
Although \sys's search algorithm is complete, the space of possible summaries to consider remains large. To address this, \sys incrementally expands the search space for program summaries to speed up the search. It does this by: (1) adding new production rules to the grammar, and (2) increasing the number of times that each product rule is expanded.

The benefits of this approach are twofold. First, since the search time for a valid summary is proportional to search space size, \sys often finds valid summaries quickly, as our experiments show. Second, since larger grammars are more syntactically expressive, the found summaries are likely to be more expensive computationally. Hence, biasing the search towards smaller grammars likely produces program summaries that run more efficiently. Although this is not sufficient to guarantee optimality of generated summaries, our experiments show that in practice \sys generates efficient solutions (\secref{perfcomparesec}).

To implement incremental grammar generation, \sys partitions the space of program summaries into different {\em grammar classes}, where each class is defined based on these syntactical features: (1) the number of MapReduce operations, (2) the number of $emit$ statements in each $map$ stage, (3) the size of key-value pairs emitted in each stage, as inferred from the types of the key and value, and (4) the length of expressions (e.g., $x+y$ is an expression of length 2, while $x+y+z$ has a length of 3). All of these features are implemented by altering production rules in the search space grammar. A grammar hierarchy is created such that all program summaries expressible in a grammar class $G_i$ are also expressible in a higher level class, i.e., $G_j$ where $j>i$. 

\subsection{\sys's Search Algorithm for Summaries}
\seclabel{sec_4.3_fullalgo}

\figref{sec_4.3_fullalgorithm} shows \sys's algorithm for searching program summaries. The algorithm begins by constructing a grammar $G$ using the results of program analysis $A$ on the input code. First, \sys partitions the grammar $G$ into a hierarchy of grammar classes $\Gamma$ (line \ref{lst:genGrammar}). Then, it incrementally searches each grammar class $\gamma \in \Gamma$, invoking the synthesizer to find summaries in $\gamma$ (line \ref{lst:callSynth}). Each summary (and invariants) the synthesizer returns is checked by a theorem prover (line ~\ref{lst:fullVerify}); \sys saves the set of correct program summaries in $\Delta$ and all summaries that fail verification in $\Omega$. Each synthesized summary (correct or not) is eliminated from the search space, forcing the synthesizer to generate a new summary each time, as explained in \secref{sec_4.1_verifierfailures}. When the grammar $\gamma$ is exhausted, i.e., the synthesizer has returned null, \sys returns the set of correct summaries $\Delta$ if it is non-empty. Otherwise, no valid solution was found, and the algorithm proceeds to search the next grammar class in $\Gamma$. If $\Delta$ is empty after exploring every grammar in $\Gamma$, i.e., no summary could be found in the entire search space, the algorithm returns null.

\subsection{Row-wise Mean Revisited}
We now illustrate how {\tt findSummary} searches for program summaries using the row-wise mean benchmark discussed in \secref{sec_2.2_example}. \figref{sec_4.2_eqClassFig} shows three sample (incremental) grammars \sys generated as a result of calling {\tt generateClasses} (\figref{sec_4.3_fullalgorithm}, line~\ref{lst:genGrammar}) along with their properties. For example, the first class, $G_1$, consists of program summaries expressed using a single $map$ or $reduce$ operator, and the transformer functions $\lambda_m$ and $\lambda_r$ are restricted to emit only one integer key-value pair. A few candidates for $\lambda_m$ are shown in the figure. For instance, the first candidate, $(i,j,v) \rightarrow [(i,j)]$, maps each matrix entry to its row and column as the output.

If {\tt findSummary} fails to find a valid summary in $G_1$ for the benchmark, it advances to the next grammar class, $G_2$. $G_2$ expands upon $G_1$ by including summaries that consist of two $map$ or $reduce$ operators, and each $\lambda_m$ can emit up to 2 key-value pairs. The search next moves to $G_3$, where $G_3$ expands upon $G_2$ with summaries that include up to three $map$ or $reduce$ operators, and the transformers can emit either integers or tuples. As shown in~\figref{rwm_java}, a valid summary is finally found in $G_3$ and added to $\Delta$. Search continues in $G_3$ for other valid summaries in the same grammar class. The search terminates after all valid summaries in $G_3$, i.e., those returned by the synthesizer and fully verified, are found. This includes the one shown in~\figref{rwm_java}.

\section{Finding Efficient Translations}
\seclabel{sec_5_optimal}
There often exist many semantically equivalent MapReduce implementations for a given sequential code fragment, with significant performance differences. Many frameworks come with optimizers that perform low-level optimizations (e.g., fusing multiple map operators). However, performing {\em semantic transformations} is often difficult.  For instance, at least three different implementations of the StringMatch benchmark exist in MapReduce, and they differ in the type of key-value pairs the $map$ stage emits (see~\secref{dyn_tuning}). Although it is difficult for a low-level optimizer to discover these equivalences by syntax analysis, \sys can perform such optimization because it searches for a high-level program summary expressed using the IR. We now discuss \sys's use of a cost model and runtime monitoring module for this purpose.

\subsection{Cost Model}
\seclabel{sec_4.4_costmodel}
\sys uses a cost model to evaluate different semantically equivalent program summaries that are found for a code fragment. Because \sys aims to translate data-intensive applications, its cost model estimates data transfer costs as opposed to compute costs.

Each synthesized program summary is a sequence of $map$, $reduce$ and $join$ operations. The semantics of these operations are known, but the transformer functions that they use ($\lambda_m$ and $\lambda_r$) are synthesized and determine the operation's cost. We define the cost functions of the {\em map}, {\em reduce} and {\em join} operations below:
%$cost(\lambda, N, W)$, where $\lambda$ is the transformer function for each $map$ or $reduce$, $N$ is the number of elements in the input data, and $W$ is a weight factor.
%
%\vspace{-0.2cm}
\begin{align}
\label{mapEqn}
cost_m(\lambda_m, N, W_m) &= W_m * N * \sum_{i=1}^{|\lambda_m|} sizeOf(emit_i) * p_i \\[0.7em]
\label{reduceEqn}
cost_r(\lambda_r, N, W_r) &= W_r * N * sizeOf(\lambda_r) * \epsilon(\lambda_r) \\[0.7em]
\label{joinEqm}
cost_j(N_1, N_2, W_j) &= W_j * N_1 * N_2 * sizeOf(emit_j) * p_j
\end{align}

The function $cost_m$ estimates the amount of data generated in the {\em map} stage. For each $emit$ statement in $\lambda_m$, the size of the key-value pair emitted is multiplied by the probability that the $emit$ statement will execute ($p_i$). The values are then summed to get the expected size of the output record. The total amount of data emitted during the map stage equals to the product of expected record size and the number of times $\lambda_m$ is executed ($N$). The cost function for a {\em reduce} stage, $cost_r$, is defined similarly, except that $\lambda_r$ produces only a single value and the cost is adjusted based on whether $\lambda_r$ is commutative and associative. The function $\epsilon$ returns 1 if these properties hold; otherwise, it returns $W_{csg}$. The cost function for {\em join} operations, $cost_j$, is defined over: the number of elements in the two input datasets ($N_1$ and $N_2$), the selectivity of the join predicate ($p_j$), and the size of the output record. $W_m$, $W_r$ and $W_j$ are the weights assigned to the map, reduce and join operations. $W_{csg}$ is the penalty for a non-commutative associative reduction. In our experiments, we used the values 1, 2, 2 and 50 for these weights, respectively based on our empirical studies.

To estimate the cost of a program summary, we simply sum the cost of each individual operation. The first operator in the pipeline takes symbolic variables $N_{0..i}$ as the number of records. For each subsequent stage, we use the number of key-value pairs generated by the current stage, expressed as a function over $N_{0..i}$:
\begin{align*}
cost_{mr}([(op_1,\lambda_1),(op_2,\lambda_2),\ldots],N_{0..i}) = cost_{op1}(\lambda_1,N,W) \; + \\
cost_{mr}([(op_2,\lambda_2),\ldots], count(\lambda_1,N_{0..i}))
\end{align*}
The function $count$ returns the number of key-value pairs generated by a given stage. For {\em map} stages, this equals $\sum_{i=1}^{|emits|} p_i$; for {\em reduce} stages, it equals the number of unique key values on which the reducer was executed; for joins, it equals $N_1 * N_2 * p_j$.

\subsection{Dynamic Cost Estimation}
\seclabel{sec_5_runtimemodule}

The cost model computes the cost of a program summary as a function of input data size $N$. We use this cost model to compare the synthesized summaries both statically and dynamically. First, calling {\tt findSummary} returns a list of verified summaries that were found. \sys then uses the cost model to prune summaries when a less costly one exists in the list. Not all summaries can be compared that way, however, since they could depend on the value distribution of the input data or how frequently a conditional evaluates to true, as shown in the candidates for grammar $G_3$'s $\lambda_{m1}$ in~\figref{sec_4.2_eqClassFig}.

In such cases, \sys generates code for all remaining summaries that have been validated, and it uses a runtime monitoring module to evaluate their costs dynamically when the generated program executes. As the program executes, the runtime module samples values from the input dataset (\sys currently uses first-k values sampling, although different sampling method may be used). It then uses the samples to estimate the probabilities of conditionals by counting the number of data elements in the sample for which the conditional will evaluate to true. Similarly, it counts the number of unique data values that are emitted as keys. These estimates are inserted into Eqn~\ref{mapEqn} and Eqn~\ref{reduceEqn} for each program summary to get comparable cost values. Finally, the summary with the lowest cost is executed at runtime. Hence, if the generated program is executed over different data distributions, it will run different implementations, as illustrated in~\secref{dyn_tuning}.

\section{Implementation}
\seclabel{sec_6_impl}
We implemented \sys using the Polyglot framework~\cite{polyglot} to parse Java code into an abstract syntax tree (AST). \sys traverses the program AST to identify candidate code fragments, performs program analysis, and generates target code. We now describe the Java features supported by our compiler front-end. We also discuss how \sys identifies code fragments for translation and generates executable code from the verified program summary.

\subsection{Supported Language Features}
\seclabel{sec_6.1_limitations}
To translate a code fragment, \sys must first successfully generate verification conditions for that fragment (as explained in \secref{sec_3.2_verifcond}). \sys can currently do this for basic Java statements, conditionals, functions, user-defined types, and loops. 

\noindent\paragraph{Basic Types} \sys supports all basic Java arithmetic, logical, and bit-wise operators. It can also process reads and writes into primitive arrays and common Java Collection interfaces, such as {\tt java.util.\{List, Set,Map\}}. \sys can be extended to support other data structures, such as {\tt Stack} or {\tt Queue}.

\noindent\paragraph{User-defined Types}
\sys traverses the program AST to find declarations of all types that were used in the code fragment being translated. It then dynamically translates and adds these types to the IR as {\tt structs}, as shown in Appendix~\ref{appendix:ir}.

\noindent\paragraph{Loops} \sys computes VCs for different types of loops ({\tt for}, {\tt while}, {\tt do}), including those with loop-carried dependencies~\cite{dragonBook}, after applying classical transformations~\cite{dragonBook} to convert loops into the {\tt while(true)\{...\}} format.

\noindent\paragraph{Methods}
\sys handles methods by inlining their bodies. Polymorphic methods can be supported similarly by inlining different versions with conditionals that check the type of the host object at runtime. Recursive methods and methods with side-effects are not currently supported because they are unlikely to gain any speedup by being translated to MapReduce.

\noindent\paragraph{External Library Methods}
\sys supports common library methods from standard Java libraries (e.g., {\tt java.lang.Math} methods) by modeling their semantics explicitly using the IR. Users can similarly provide models for other methods that \sys currently does not support.~\footnote{We provide examples of library function and type models in Appendix~\ref{appendix:ir}.}

\subsection{Code Fragment Identification}
\sys traverses the input AST to identify code fragments that are amenable for translation by searching for loops that iterate one or more data structures (e.g., a list or an array). We target loops since they are most likely to benefit from translation to MapReduce. We have kept our loop selection criteria lenient to avoid false negatives.

\subsection{Code Generation}
\seclabel{sec_6.1_codegen}
Once an identified code fragment is translated, \sys replaces the original code fragment with the translated MapReduce code. It also generates ``glue'' code to merge the generated code into the rest of the program. This includes creating a {\tt SparkContext} (or an {\tt ExecutionEnvironment} for Flink), converting data into {\tt RDD}s (or Flink's {\tt DataSet}s), broadcasting required variables, etc. Since some API calls (such as Spark's {\tt reduceByKey}) are not defined for non-commutative associative transformer functions, \sys uses these API calls only if the generated code is indeed commutative and associative (otherwise, \sys uses safe, albiet less efficient, transformations, such as {\tt groupByKey}). Finally, \sys also generates code for sampling input data and dynamic switching, as discussed in \secref{sec_5_runtimemodule}. Appendix~\ref{appendix:codegen} presents a subset of code-generation rules for the Spark API.

\section{Evaluation}
\seclabel{evaluation}
In this section, we present a comprehensive evaluation of \sys on a number of dimensions, including its ability to: (1) handle diverse and realistic workloads, (2) find efficient translations, (3) compile efficiently, and (4) extend to support other IRs and cost-models in the future. All experiments were conducted on an AWS cluster of 10 m3.2xlarge instances (1 master node, 9 core nodes), where each node contains an Intel Xeon 2.5 GHz processor with 8 vCPUs, 30 GB of memory, and 160 GB of SSD storage. We used the latest versions of all frameworks available on AWS: Spark 2.3.0, Hadoop 2.8.3, and Flink 1.4.0. The data files for all experiments were stored on HDFS.

\begin{table}
\centering
\small
\begin{tabular}[t]{ |l|c|c|c| }
 \hline
 \textbf{Suite}	&	\textbf{\# Translated}	&	\textbf{Mean Speedup} &	\textbf{Max Speedup} 	\\
 \hline
 Phoenix  				& 7 / 11	& 	14.8x 	&  32x	\\
\hline	
 Ariths					& 11 / 11	& 	12.6x	&  18.1x	\\
\hline
 Stats					& 18 / 19	&  	18.2x  	&  28.9x	\\
\hline
 \biglam				& 6 / 8		&  	21.5x   &  32.2x	\\
\hline
 Fiji					& 23 / 35	&   18.1x 	&  24.3x   	\\
\hline
%TPC-H					& 10 / 10   &   1.3x 	&  2.8x   	\\
 TPC-H					& 10 / 10   &   31.8x 	&  48.2x   	\\
\hline
% Iterative				& 7 / 7	    &   0.8x 	&  1x   	\\
 Iterative				& 7 / 7	    &   18.4x 	&  28.8x   	\\
\hline
\end{tabular}
\vspace{0.2cm}
\caption{Number of code fragments translated by \sys and their mean and max speedups compared to sequential implementations.}
\tablabel{feasibility}
\vspace{-0.3in}
\end{table}

% Contains External Library Calls	 	& \maaz{xx}							\\
%\hline
% Average Lines of Code    	 		 & xx    \\
%\hline 
%\hline
% \textbf{Output Code Properties}     &  \textbf{Number of Benchmarks}    \\
% \hline 
% Contains Conditionals               & 11   \\
%\hline  
% Contains Tuples                     & 6    \\
%\hline  
% Multiple Output Variables           & 5    \\
%\hline
% 1 MapReduce operation               & 17   \\
%\hline
% 2 MapReduce operations              & 31   \\
%\hline
% 3 MapReduce operations              & 2    \\
%\hline
% Average Lines of Code    	 		 & xx    \\
%\hline

\subsection{\textbf{Feasibility Analysis}}
We first assess \sys's ability to handle a variety of data-processing applications. Specifically, we determine whether: (1) \sys can generate verification conditions for a syntactically diverse set of programs, (2) our IR can express summaries for a broad range of data-processing workloads, and (3) \sys's ability to find such summaries. To this end, we used \sys to optimize a set of \totalbench diverse benchmarks from real-world applications that contained a total of \totalcf translatable code fragments.

\paragraph{Basic Applications}
For benchmarking, we assembled a set of small applications from prior work and online repositories. These applications, summarized below, contain a diverse set of code patterns commonly found in data-processing workloads (e.g., aggregations, selections, grouping, etc), as follows:
\begin{asparaitem}
\item {\em {\biglam}}~\cite{biglambda} consists of several data analysis tasks such as {\em sentiment analysis}, {\em database operations} (e.g., selection and projection), and {\em Wikipedia log processing}. Since \biglam generates code from input-output examples rather than from an actual implementation, we recruited computer science graduate students in our department to implement a representative subset of the benchmarks from their textual descriptions. This resulted in 211 lines of code across 7 files.
\item {\em Stats} is a set of benchmarks \sys automatically extracted from an online repository for the statistical analysis of data~\cite{magpie}. Examples include {\em Covariance}, {\em Standard Error} and {\em Hadamard Product}. The repository contains 1162 lines of code across 12 Java files, mostly consisting of vector and matrix operations.
\item {\em Ariths} is a set of simple mathematical functions and aggregations collected from prior work~\cite{arith1,arith2,arith3,grassp}. Examples include {\em Min}, {\em Max}, {\em Delta}, and {\em Conditional Sum}. The suite contains 245 lines of code than span 11 files.
\end{asparaitem}

Across the 3 suites, \sys identified 38 code fragments, of which 35 were successfully translated. One code-fragment that \sys failed to translate used a variable-sized kernel to convolve a matrix; two others required broadcasting data values to many reducers during the map stage, but such mappers are currently inexpressible in our IR due to the absence of loops.

\paragraph{Traditional Data-Processing Benchmarks}
Next, we used \sys to translate a set of well-known, data-processing benchmarks that resemble real-world workloads:
\begin{asparaitem}
\item We manually implemented Q1, Q6, Q15 and Q17 from the {\em TPC-H} benchmark using sequential Java and used \sys to translate the Java implementations to MapReduce. The selected queries cover many SQL features, such as aggregations, joins and nested queries.
\item {\em Phoenix}~\cite{pheonix} is a collection of standard MapReduce problems---such as {\em 3D Histogram}, {\em Linear Regression}, {\em KMeans}, etc.---used in prior work~\cite{mold}. Since the original sequential implementations were written in C, we used the sequential Java translations of the benchmarks from prior work in our experiments. The suite consists of 440 lines of code across 7 files.
\item {\em Iterative} represents two popular iterative algorithms that we manually implemented into sequential versions: {\em PageRank} and {\em Logistic Regression Based Classification}.
\end{asparaitem}

\sys successfully translated all 4 TPC-H queries and both iterative algorithms. It successfully translated 7 of 11 from the Phoenix suite. Three of the 4 failures were due to the IR's lack of support for loops inside transformer functions. One benchmark failed to synthesize within 90 minutes, causing \sys to time out.

\paragraph{Real-World Applications}
Fiji~\cite{fiji} is a popular distribution of the ImageJ~\cite{ImageJ} library for scientific image analysis. We ran \sys on the source code of four Fiji packages (aka plugins). {\em NL Means} is a plugin for denoising images via the non-local-means algorithm~\cite{denoising} with optimizations~\cite{cryomicroscopy}. {\em Red To Magenta} transforms images by changing red pixels to magenta. {\em Temporal Median} is a probabilistic filter for extracting foreground objects from a sequence of images. {\em Trails} averages pixel intensities over a time window in an image sequence. These packages, authored by different developers, contain 1411 lines of code that span 5 files. Of the 35 candidate code fragments identified across all 4 packages, \sys successfully optimized 23. Three of the failures were caused by the use of unsupported types or methods from the ImageJ library since we did not model them using the IR, and the search timed out for the remaining 9.

\medskip

\tabref{feasibility} summarizes the results of our feasibility analysis. Of the 101 individual code fragments identified by the compiler across all benchmarks, \sys translated 82. We manually inspected all code files to ensure that \sys's code fragment identifier missed no translatable code fragments. Overall, the benchmarks form a syntactically diverse set of applications.\footnote{We summarize the syntactic features of the code fragments in Appendix~\ref{appendix:features}.}

Because MOLD is not publicly available, we obtained the generated code from the MOLD authors for the benchmarks used in its evaluation~\cite{mold}. Of the 7 Phoenix benchmarks, MOLD could not translate 2 ({\em PCA} and {\em KMeans}). Another 2 ({\em Histogram} and {\em Matrix Multiplication}) generated semantically correct translations that worked well for multi-core execution but failed to execute on the cluster because they ran out of memory. For the remaining 3 benchmarks ({\em Word Count}, {\em String Match} and {\em Linear Regression}), MOLD generated working implementations. In contrast, \sys translated 4 of the 7 Phoenix benchmarks. For {\em PCA} and {\em KMeans}, \sys translated and successfully executed a subset of all the loops found, while translation failed for the other loops and the {\em Matrix Multiplication benchmark} for reasons explained above. 

\subsection{Performance of the  Translated Benchmarks}
\seclabel{perfcomparesec}

\begin{figure}[t]
    \begin{subfigure}{\linewidth}
    \includegraphics[width=1\linewidth]{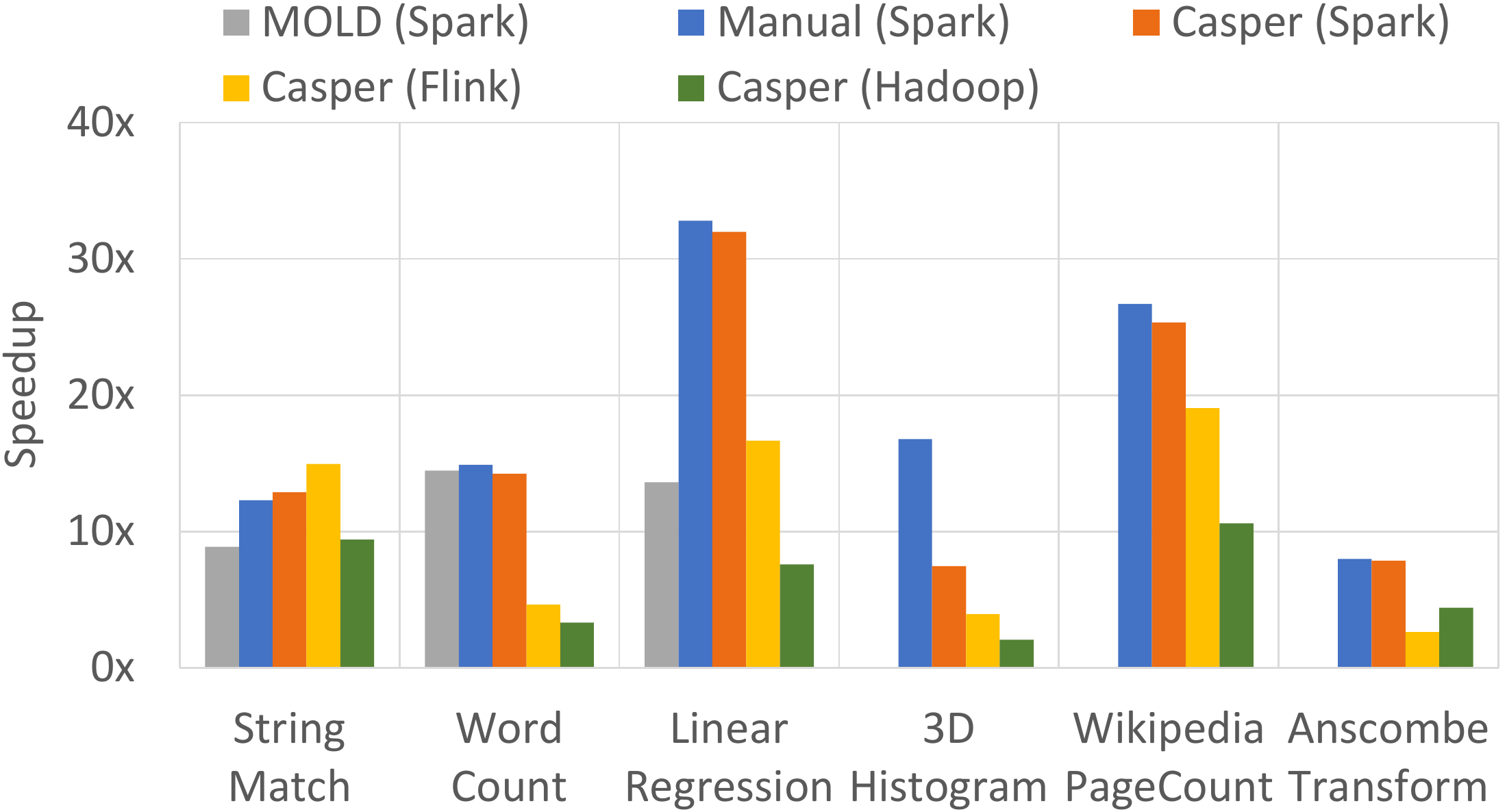}
    \vspace{-0.4cm}
    \caption{\sys achieves speedup competitive with manual translations}
    \figlabel{comparisons}
    \vspace{0.5cm}
    \end{subfigure}

    \begin{subfigure}{0.49\linewidth}
    \includegraphics[width=1\linewidth]{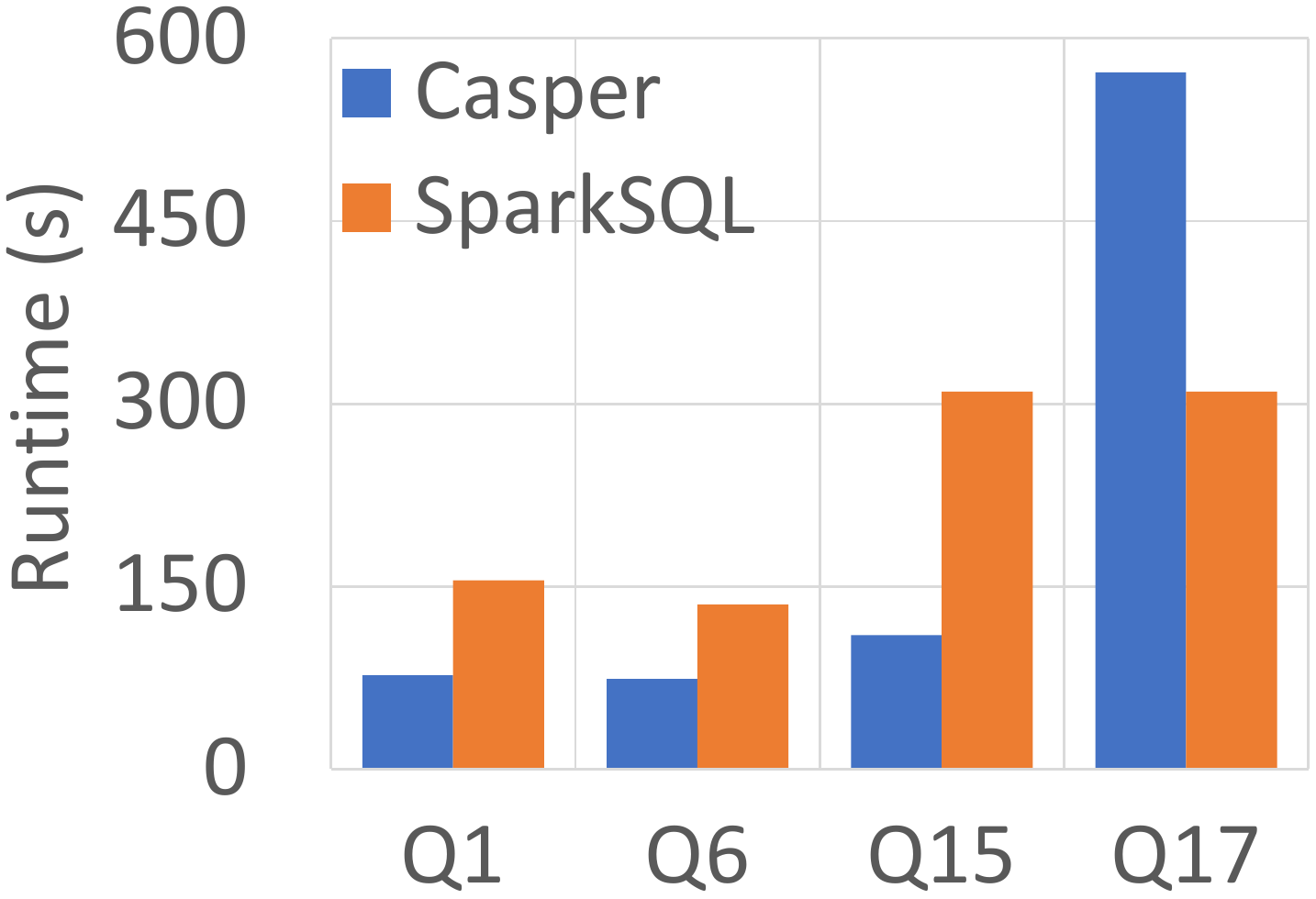}
    \caption{TPC-H benchmarks}
    \figlabel{tpch}
    \end{subfigure}
    \hspace{0.1cm}
    \begin{subfigure}{0.40\linewidth}
    \includegraphics[width=1\linewidth]{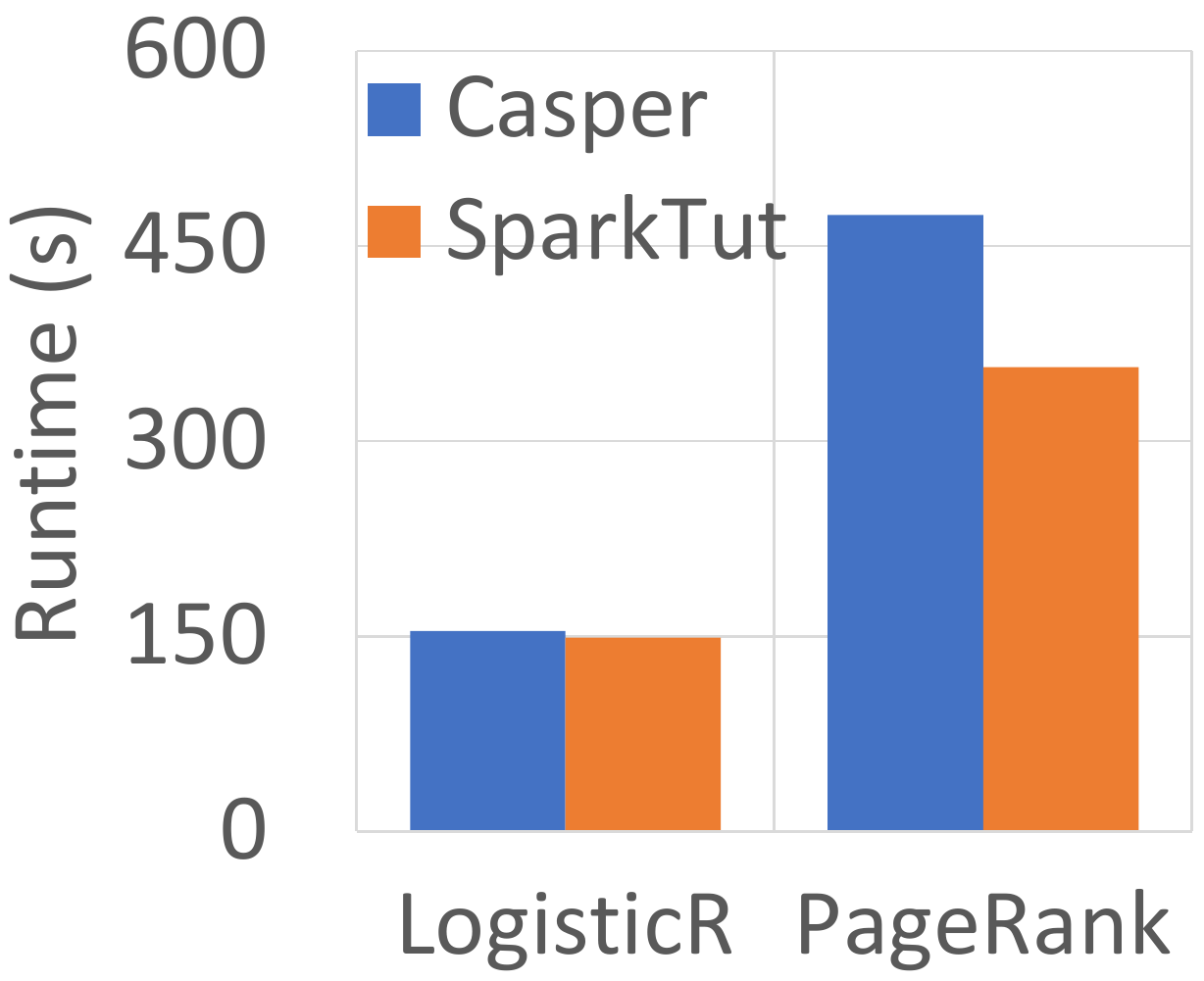}
    \caption{Iterative algorithms}
    \figlabel{iterative}
    \end{subfigure}
    \vspace{-0.1cm}
    \caption{A runtime comparison of \sys-generated implementations against reference implementations.}
    \vspace{-0.4cm}
    \figlabel{test}
\end{figure}

\sys helps an application leverage the optimization and parallelization provided by MapReduce implementations by translating their code. Therefore, in this section, we examine the quality of the translations \sys produced by comparing their performance to that of reference distributed implementations.

We used \sys to translate summaries for these benchmarks to three popular implementations of the MapReduce programming model: Hadoop, Spark, and Flink. The translated Spark implementations, along with their original sequential implementations, were executed on three synthetic datasets of sizes 25GB, 50GB, and 75GB. Overall, the Spark implementations \sys generated are $15.6\times$ faster on average than their sequential counterparts, with a max improvement of up to $48.2\times$. \tabref{feasibility} shows the mean and max speedup observed for each benchmark suite using Spark on a 75GB dataset. We also executed the Hadoop and Flink implementations generated by \sys for a subset of 10 benchmarks, some of which are shown in \figref{comparisons}. The average speedups observed (over the 10 benchmarks) by these implementations are 6.4$\times$ and 10.8$\times$, respectively. These results show that \sys can effectively improve the performance of applications by an order of magnitude by retargeting critical code fragments for execution on MapReduce frameworks.

\figref{comparisons} plots the speedup achieved by the MOLD-generated implementations for {\em String Match}, {\em Word Count}, and {\em Linear Regression}. The Spark translations MOLD generated for these benchmarks performed 12.3$\times$ faster on average than the sequential versions. The solutions generated by \sys for {\em String Match} and {\em Linear Regression} were faster than those generated by MOLD by 1.44$\times$ and 2.34$\times$, respectively. For {\em String Match}, \sys found an efficient encoding to reduce the amount of data emitted in the map stage (see~\secref{dyn_tuning}), whereas MOLD emitted a key-value pair for every word in the dataset. Furthermore, MOLD used separate MapReduce operations to compute the result for each keyword; \sys computed the result for all keywords in the same set of operations. For {\em Linear Regression}, MOLD discovered the same overall algorithm as \sys except its implementation zipped the input RDD with its index as a pre-processing step, almost doubling the size of input data and hence the amount of time spent in data transfers. 

For the {\em Ariths}, {\em Stats}, {\em Big$\lambda$}, and {\em Fiji} benchmarks, we recruited Spark developers through UpWork.com to manually rewrite the benchmarks since reference distributed implementations were not available.\footnote{Appendix~\ref{appendix:developer} describes the hiring criteria.} \figref{comparisons} compares the performance of (a subset of) \sys-generated implementations to handwritten benchmark implementations over the 75GB dataset. Results show that the \sys-generated implementations perform competitively, even with those manually written by developers. In fact, of the 42 hand-translated benchmark implementations, 24 used the same high-level algorithm as the one generated by \sys, and most of the remaining ones differ by using framework-specific methods instead of an explicit map/reduce (e.g., using Spark's built-in filter, sum, and count methods). However, these variations did not cause a noticeable performance difference. One interesting case was the 3D Histogram benchmark, where the developer exploited knowledge about the data to improve runtime performance. Specifically, the developer recognized that since RGB values always range between 0-255, the histogram data structure would never exceed 768 values. Therefore, the developer used Spark's more efficient $aggregate$ operator to implement the solution. \sys, not knowing that pixel RGB values are bounded, assumed that the number of keys could grow to be arbitrarily large and that using the aggregate operator could cause out-of-memory errors, hence it generated a single stage map and reduce instead.

For {\em PageRank} and {\em Logistical Regression}, we compared \sys against the implementations found in the Spark Tutorials~\cite{sparktuts} (see \figref{iterative}). The reference PageRank implementation was 1.3$\times$ faster than the one \sys generated on a dataset of about 2.25 billion graph edges and running 10 iterations. This is because \sys currently does not generate any {\tt cache()} statements, nor does it  co-partition data. Deciding when to cache can lead to further performance gains. Prior work~\cite{systemml} suggested heuristics for inserting such statements into Spark algorithms that could be integrated into \sys's code generator to improve performance for iterative workloads. For {\em Logistical Regression}, we found no noticeable difference in performance.

For TPC-H queries, we compared the performance of Spark code generated by \sys against SparkSQL's implementation. \figref{tpch} plots the results of this experiment. For Q1, Q6 and Q15, \sys implementations executed 2$\times$, 1.8$\times$ and 2.8$\times$ faster, respectively, than SparkSQL on a scale factor of 100. For Q1 and Q6, we attribute this to the extra data shuffling performed by the SparkSQL query plan. In Q15, SparkSQL's query plan scanned the {\em lineitem} relation twice, whereas \sys's implementation did so only once, resulting in worse runtime performance. For Q17, SparkSQL executed 1.7$\times$ faster because it performed better scheduling of the query operators than the \sys-generated implementation. In sum, results show that the \sys-generated implementations the TPC-H benchmarks have comparable performance to those implemented directly using the MapReduce frameworks. Yet, developers need not learn different MapReduce APIs by using \sys.

\subsection{Compilation Performance}
\seclabel{compileperf}
We next evaluate \sys's compilation performance. We discuss the time taken by \sys to compile the benchmarks, the effectiveness of \sys's two-phase verification strategy, the quality of the generated code, and incremental grammar generation.

\subsubsection{Compile Time} On average, \sys took 11.4 minutes to compile a single code fragment. However, the median compile time for a single benchmark was only 2.1 minutes: for some benchmarks, the synthesizer discovered a low-cost solution during the first few grammar classes, letting \sys terminate search early. \tabref{compilation} shows the mean compilation time for a single benchmark by suite.

\begin{table}
\small
\centering
\begin{tabular}{ |Q{4em}|P{4em}|P{4em}|P{4em}|P{4em}| }
 \hline
 	\textbf{Source}		&	
 	\textbf{Mean Time (s)}	&	
 	\textbf{Mean LOC} 	&	
 	\textbf{Mean \# Op} 	&
 	\textbf{Mean TP Failures}		\\
 \hline
 Phoenix  	& 944	& 	13.8 (13.1)	&  2.3 (2.1) & 0.35	\\
\hline	
 Ariths		& 223	& 	9.4 (7.6)	&  1.6 (1.2) & 4	\\
\hline
 Stats		& 351	&  	7.6 (5.8)  	&  1.8 (1.8) & 0.6	\\
\hline
 \biglam	& 112	&  	13.6 (10)    &  1.8 (2.0) & 0.4	\\
\hline
 Fiji		& 1294	&   7.2 (7.4) 	&  1.4 (1.6) & 0.1	\\
\hline
 TPC-H		& 476	    &  5.9 (n/a) &  7.25 (n/a)  & 0	\\
\hline
 Iterative	& 788	    &  3.3 (3.7) &  4.5 (3.5)  & 2	\\
\hline
\end{tabular}

\caption{Summary of \sys's compilation performance. Values for the reference implementations are shown in parentheses.}
\tablabel{compilation}
\vspace{-0.70cm}
\end{table}

\subsubsection{Two-Phase Verification}
\seclabel{twophaseverifeval}
In our experiments, the candidate summary generator produced at least one incorrect solution for 13 out of the \totalcf successfully translated code-fragments. The synthesizer proposed a total of 76 incorrect summaries across all benchmarks. \tabref{compilation} lists the average number of times the theorem prover rejected a solution for each benchmark suite. As an example, the {\em Delta} benchmark computes the difference between the largest and smallest values in the dataset. It incurred 7 rounds of interaction with the theorem prover before the candidate generator found a correct solution due to errors from bounded model checking (discussed in~\secref{sec_4.1_verifierfailures}).

\subsubsection{Generated Code Quality} In addition to measuring the runtime performance of \sys-generated implementations, we manually inspected the code generated by \sys and compared it to the reference implementations for two code quality metrics: lines of code (LOC) and the number of MapReduce operations used. \tabref{compilation} shows the results of our analysis. Implementations generated by \sys were comparable and did not use more MapReduce operations or LOC than were necessary to implement a given task. Note that the LOC pertain to individual code fragments, not entire benchmarks.

\begin{table}
\centering
\small
\begin{tabular}{ |l|P{7em}|P{7em}| }
 \hline
 \textbf{Benchmark} &   \textbf{With Incr. Grammar} & \textbf{Without Incr. Grammar} \\
\hline
 WordCount    			&  2 & 827 \\
\hline  
 StringMatch  			&  24& 416 \\
\hline  
 Linear Regression  	&  1 & 94  \\
\hline  
 3D Histogram  			&  5 & 118 \\
\hline  
 YelpKids  				&  1 & 286 \\
\hline  
 Wikipedia PageCount  	&  1 & 568 \\
\hline  
 Covariance  			&  5 & 11  \\
\hline  
 Hadamard Product  		&  1 & 484 \\
\hline  
 Database Select  		&  1 & 397 \\
\hline  
 Anscombe Transform  	&  2 & 78  \\
\hline
\end{tabular}

\caption{With incremental grammar generation, \sys produces far less redundant summaries.}
\tablabel{pruning}
\vspace{-0.7cm}
\end{table}

\subsubsection{Incremental Grammar Generation} 
\seclabel{pruningeffect}
We also measured the effectiveness of incremental grammar generation in optimizing search. To measure its impact on compilation time, we used \sys to translate benchmarks without incremental grammar generation and compared the results. The synthesizer was allowed to run for 90 minutes, after which it was manually killed. The results of this experiment are summarized in \tabref{pruning}. Exhaustively searching the entire search space produced hundreds of more expensive solutions. The cost of searching, verifying, and sorting all these superfluous solutions dramatically increased overall synthesis time. In fact, \sys timed out for every benchmark in that set (which represents a slowdown by at least one order of magnitude). 

\begin{figure*}[t]
\centering
\begin{minipage}[t]{0.27\linewidth}
\centering
\begin{subfigure}{1\linewidth}
\begin{lstlisting}[language=psuedo]
key1_found = false
key2_found = false
for word in text:
  if word == key1:
    key1_found = true;
  if word == key2:
    key2_found = true;
\end{lstlisting}
\caption{Sequential code for StringMatch}
\figlabel{dynamic_orig}
\end{subfigure}
\end{minipage}
\begin{minipage}[t]{0.39\linewidth}
\centering
\hspace{0.1cm}
\begin{subfigure}{0.7\linewidth}
\includegraphics[width=1\linewidth]{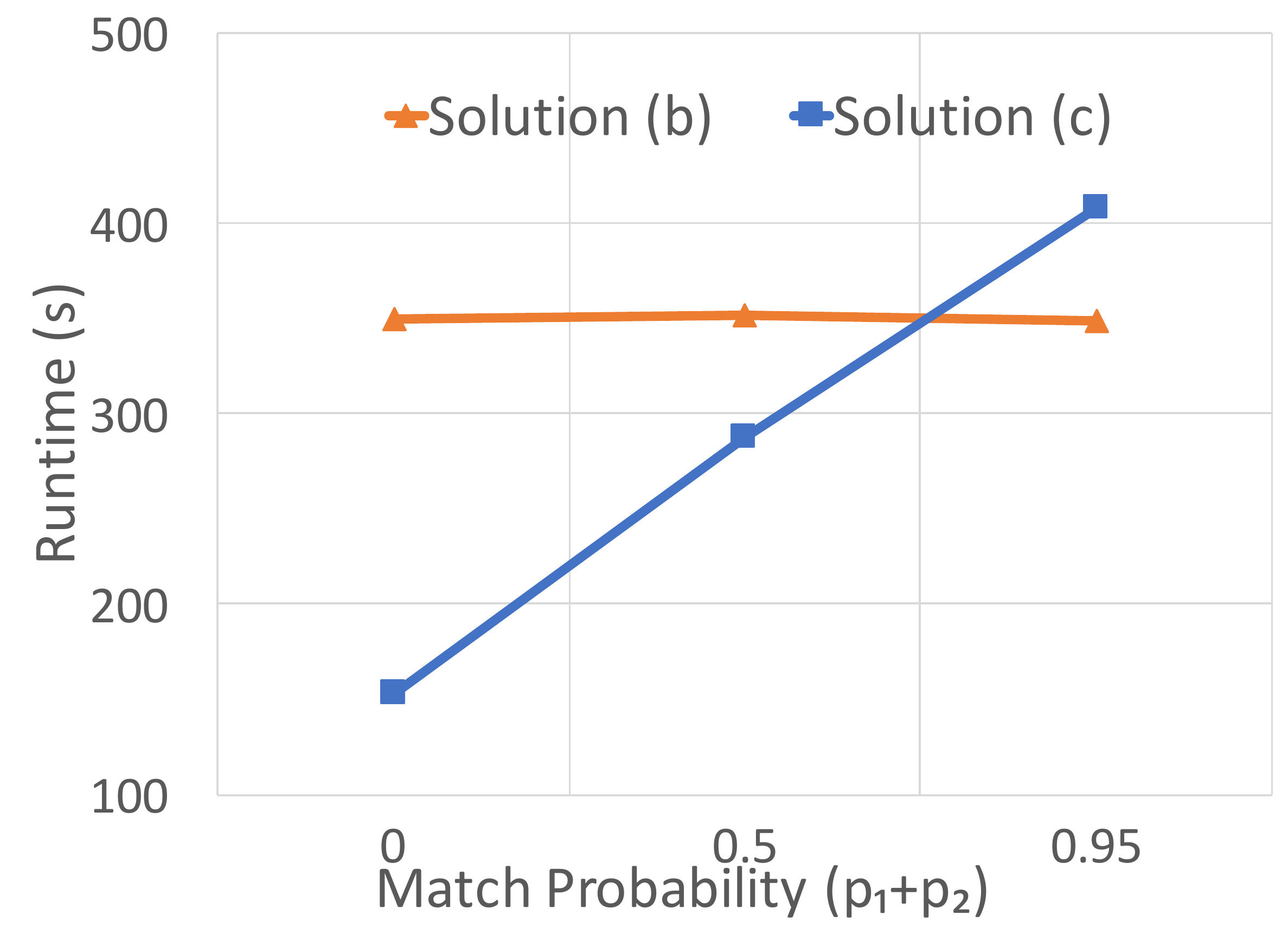}
\caption{Performance of solutions over datasets with different levels of skew}
\vspace{0.5cm}
\figlabel{dynamic_res}
\end{subfigure}
\end{minipage}
\begin{minipage}[b]{0.31\linewidth}
\centering

\begin{subfigure}{\linewidth}
\vspace{0.90cm}
\hspace{0.2cm}
\begin{small}
\begin{tabular}{ |l|P{4em}|P{4em}| }
\hline
 \textbf{Dataset}  &  \textbf{Cost of Soln (c)} & \textbf{Optimal Solution}  \\
\hline
  0\% match & 0 & (c)  \\
\hline
  50\% match & 75$N$ & (c) \\
\hline
  95\% match & 142.5$N$ & (b) \\
\hline
\end{tabular}
\end{small}
\vspace{0.55cm}
\caption{Dynamic selection of optimal algorithm}
\vspace{0.5cm}
\figlabel{dynamic_fcosts}
\end{subfigure}
\end{minipage}
\vspace{-0.17in}

\begin{subfigure}[t]{\linewidth}
\begin{center}
\begin{footnotesize}
\begin{tabular}{ |l|l|l| }
\hline
 & \textbf{Solution}  &   \textbf{Static Cost}  \\
\hline
  a &
  \pbox{20cm}{
      \vspace{0.2cm}
      $output = reduceByKey(map(text,\lambda_{m}),\lambda_{r})$ \\
      $\lambda_{m} : (word) \rightarrow \{ (key1,  word=key1), (key2,  word = key2)\}$ \\
      $\lambda_{r} : (v_1, v_2) \rightarrow v_1 \, \vee \; v_2$
      \vspace{0.2cm}
  }
  &   
  \pbox{20cm}{
      $\lambda_m : 2*(40+10)*N$ \\
      $\lambda_r : 2*2*50*N$ \\
      Total : $300N$ 
  }
  \\
\hline
  b &
  \pbox{20cm}{
      \vspace{0.2cm}
      $output = reduce(map(text,\lambda_{m}),\lambda_{r})$ \\
      $\lambda_{m} : (word) \rightarrow \{ (word=key1, \; word=key2) \}$ \\
      $\lambda_{r} : (t_1, t_2) \rightarrow (t_1[0] \; \vee \; t_2[0], \; t_1[1] \; \vee \; t_2[1])$
      \vspace{0.2cm}
  }
  &
  \pbox{20cm}{
      $\lambda_m : 1*28*N$ \\
      $\lambda_r : 2*28*N$ \\
      Total : $84N$ 
  }
  \\
\hline
  c &
  \pbox{20cm}{
      \vspace{0.2cm}
      $output = reduceByKey(map(text,\lambda_{m}),\lambda_{r})$ \\
      $\lambda_{m} : (word) \rightarrow \{ if \: (word = key1): (key1, true), 
                                           if \: (word=key2): (key2, true)\}$ \\
      $\lambda_{r} : (v_1, v_2) \rightarrow v_1 \, \vee \; v_2$
      \vspace{0.2cm}
  }
  &   
  \pbox{20cm}{
      $\lambda_m : (p_1+p_2)*50*N$ \\
      $\lambda_r : (p_1+p_2)*2*50*N$ \\
      Total : $150(p_1+p_2)$ 
  }
  \\
\hline
\end{tabular}
\end{footnotesize}
\end{center}
\vspace{-0.15cm}
\caption{Candidate solutions and their statically computed costs}
\figlabel{dynamic_sols}
\end{subfigure}
\vspace{-0.35cm}
\caption{StringMatch benchmark: \sys dynamically selects the optimal implementation
for execution at runtime.}
\figlabel{dynamic}
\vspace{-0.1cm}
\end{figure*}

\subsection{Dynamic Tuning}
\seclabel{dyn_tuning}
The final set of experiments evaluated the runtime monitor module and whether the dynamic cost model could select the correct implementations. As explained in \secref{sec_5_runtimemodule}, the performance of some solutions depends on the distribution of the input data. Therefore, we used \sys to generate different implementations for the StringMatch benchmark (\figref{dynamic_orig}). \figref{dynamic_sols} shows three (out of 400+) correct candidate solutions, with their respective costs based on the formula described in \secref{sec_4.4_costmodel} and the following values for data-type sizes: 40 bytes for String, 10 bytes for Boolean and 28 bytes for a tuple of Boolean Objects. Solution (a) can be disqualified at compile time because it will have a higher cost than solution (b) for all possible data distributions. However, the cost of solutions (b) and (c) cannot be statically compared due to the unknowns $p_1$ and $p_2$ (the respective probabilities that the conditionals will evaluate to true and a key-value pair will be emitted). The values of $p_1$ and $p_2$ depend on the input data, i.e., how often the keywords appear in the text, and thus can be determined only dynamically at run-time.

\sys handles this by generating a runtime monitor in the output code. The monitor samples the input data (first 5000 values) in each execution to estimate values for unknown variables in the cost formulas. The estimated values are then plugged back into the original cost functions (Eqn~\ref{mapEqn} and~\ref{reduceEqn}), and the solution with the lowest cost is then executed.

We executed solutions (b) and (c) on three 75GB datasets with different amounts of skew: one with no matching words (i.e., (c) emits nothing), one with 50\% matching words (i.e., (c) emits a key-value pair for half of the words in the dataset), and one with 95\% matching words (i.e., (c) emits a key-value pair for 95\% of the words in the dataset). \figref{dynamic_fcosts} shows the dynamically computed final cost of solution (c) using $p_1$ and $p_2$ estimates calculated using sampling. \figref{dynamic_res} shows the actual performance of the two solutions. For datasets with very high skew, it is beneficial to use solution (b) due to the smaller size of its key-value pair emit. Otherwise, solution (c) performs better. \sys, with the help of the dynamic input from the runtime monitor, makes this inference and selects the correct solution for all three datasets.

Dynamic cost estimation is particularly impactful in workloads with multiple join operations. The size of each relation participating in the join in addition to the selectivity of the join predicate dictate the most cost-efficient join ordering. To demonstrate this, we translated a simple query based on the TPC-H schema that implements a 3-way join between the {\em part}, {\em supplier}, and {\em partsupplier} relations. Query parameters are the name of the supplier and the customer\_id, and outputs are the customer's name, email address, and the sum of discount savings across all sales between the two parties. We executed this query over two parameter configurations: one where the cardinality of {\em join(sales, supplier)} was much greater than {\em join(sales, customer)} and one where it was much smaller. On compilation, \sys generated two semantically equivalent implementations for the query with different join orderings; which one to use depends on the cardinality of the input data. Upon execution, the \sys runtime estimated the cost of each join ordering and executed the faster solution for both configurations, showing the effectiveness of our dynamic tuning approach. We discuss the accuracy of the cost-functions we used in Appendix~\ref{appendix:costmodel} 

\subsection{System Extensibility}
The translation techniques \sys uses are not coupled to our IR or the target frameworks used. To demonstrate \sys's extensibility, we implemented the Fold-IR in prior work~\cite{fir} in our system. Adding the {\tt fold} construct to our IR required just 5 lines of code. An additional 43 lines of code were required to implement compilation of the fold operator to Dafny for verification of synthesized summaries. Since operations such as {\tt min}, {\tt max}, {\tt set.insert} and {\tt list.append} were already available in our IR, hence no extra work was needed. We did not implement any incremental grammar exploration for Fold-IR and used a constant bound to restrict the maximum size of summary expressions. With this minimal amount of work, we synthesized summaries expressed in Fold-IR for all benchmarks in the {\em Ariths} set. We believe it should be easy to extend \sys's code generator to output the same code as in the original work.

We also explored using WeldIR~\cite{weld} to express summaries. Although WeldIR is an excellent abstraction for data-processing workloads, we believe it is not suited for synthesis because it is too low-level. However, since both our IR and Fold-IR are conceptually subsets of WeldIR, summaries expressed using them can be translated to Weld through simple rewrite rules. To demonstrate, we successfully translated the summary for TPC-H Q6 expressed in our IR to Weld and used the Weld compiler to produce vectorized, multi-threaded code. 

\section{Related Work}
\seclabel{relatedWork}

\paragraph{Implementations of MapReduce}
MapReduce~\cite{mapreduce} is a popular programming model that has been implemented by various systems~\cite{hive,spark,pig}. These systems provide their own high-level DSLs that developers must use to express their computation. In contrast, \sys works with native Java programs and infers rewrites automatically.

\paragraph{Source-to-Source Compilers}
Many efforts translate programs from low-level languages into high-level DSLs. MOLD~\cite{mold}, a source-to-source compiler, relies on syntax-directed rules to convert native Java programs to Apache Spark. Unlike MOLD, \sys translates based on program semantics and eliminates the need for rewrite rules, which are difficult to devise and brittle to code pattern changes. Many source-to-source compilers have been built similarly for other domains~\cite{bones}. Unlike prior approaches in automatic parallelization~\cite{suif, polaris}, \sys targets data parallel processing frameworks and translates only code fragments that are expressible in the IR for program summaries.

\paragraph{Synthesizing Efficient Implementations}
Prior work used synthesis to generate efficient implementations and optimize programs. \cite{biglambda} synthesizes MapReduce solutions from user-provided input and output examples. QBS \cite{qbs,summaryArticle,survey} and STNG \cite{stng} both use synthesis to convert low-level languages to specialized high-level DSLs for database applications and stencil computations, respectively. \sys takes inspiration from prior approaches by applying verified lifting to construct compilers. Unlike prior work, however, \sys: (1) addresses the problem of verifier failures and designs a grammar hierarchy to prune away non-performant summaries, (2) has a dynamic cost model and runtime monitoring module for adaptively choosing from different implementations at runtime.

\paragraph{Query Optimizers and IRs}
Modern frameworks usually ship with sophisticated query optimizers~\cite{emma,hyper,tupleware,legobase,sparksql}
for generating efficient execution plans. However, these tools make users express their queries in the provided APIs. Our objective is orthogonal, i.e., to find the best way to express program semantics using the APIs provided by these tools. We essentially enable these tools to optimize code {\em not} written in their API. Furthermore, unlike our IR, most IRs meant to capture data-processing workloads~\cite{fir,weld} are not designed with synthesis in mind. This makes it difficult both to find and verify programs expressed in them.

\section{Conclusion}
\seclabel{conclusion}
We presented \sys, a new compiler that identifies and converts sequential Java code fragments into MapReduce frameworks. Rather than defining pattern-matching rules to search for convertible code fragments, \sys instead automatically discovers high-level summaries of each input code fragment using program synthesis and retargets the found summary to the framework's API. Our experiments show that \sys can convert a wide variety of benchmarks from both prior work and real-world applications and can generate code for three different MapReduce frameworks. The generated code performs up to 48.2$\times$ faster compared to the original implementation, and is competitive with translations done manually by developers.

\section{Acknowledgements}
This work is supported by the National Science Foundation through grants IIS-1546083, IIS-1651489, OAC-1739419, and CNS-1563788; DARPA award FA8750-16-2-0032; DOE award DE-SC0016260; the Intel-NSF CAPA center, and gifts from Adobe, Amazon, and Google.

\bibliographystyle{ACM-Reference-Format}
\bibliography{paper}

\appendix
\vfill\eject
\section{Proof Sketch For Soundness and Completeness}
\label{appendix:proof}
Here, we first formalize the definitions of soundness and completeness, and then we present a proof sketch to show that \sys's synthesis algorithm for program summaries has these properties. We use terms and acronyms defined in the paper without explaining them again here.
\\[0.8em]
\textbf{Definition 1.} {\em (Soundness of Search)}~~ An algorithm for generating program summaries is sound if and only if, for all program summary $ps$ and loop invariants $inv_1,\ldots,inv_n$ generated by the algorithm, the verification conditions hold over all possible program states after we execute the input code fragment $P$. In other words, $\forall \sigma. \; VC(P, ps, inv_1,\ldots,inv_n, \sigma)$.
\\[0.8em]
\textbf{Definition 2.} {\em (Completeness of Search)}~~An algorithm for generating program summaries is complete if and only if when there exists $ps, inv_1,\ldots,inv_n\in G$, then $\forall \sigma. \; VC(P, ps, inv_1,\ldots,inv_n, \sigma)) \rightarrow (\Delta \ne \emptyset)$. Here, $G$ is the search space traversed, $P$ is the input code fragment, $VC$ is the set of verification conditions, and $\Delta$ is the set of sound summaries found by the algorithm. In other words, the algorithm will never fail to find a correct program summary as long as one exists in the search space.
%\\[0.8em]
\paragraph{\textbf{Proof of Soundness.}} The soundness guarantee for \sys's synthesis algorithm is derived from the soundness guarantees offered by Hoare-style verification conditions. The proof is constructed using a {\em loop-invariant}, namely, a statement that is true immediately before and after each loop execution. Hoare logic dictates that in order to prove correctness of a given postcondition (i.e., program summary) for a given loop, we must prove the following holds over all possible program states:

\begin{enumerate}
\item The invariant is true before the loop.
\item Each iteration of the loop maintains the invariant.
\item Once the loop has terminated, the invariant implies the postcondition.
\end{enumerate}

This is essentially an inductive proof. The first two constraints prevent \sys from finding a loop invariant strong enough to imply an incorrect program summary. Our correctness guarantee is, of course, subject to the correct implementation of our VC generation module and of the theorem prover we use (Dafny). Establishing that the summary is a correct postcondition is sufficient to establish that it is a correct translation. This is so because summaries in our IR must describe the final value of {\em all} output variables (i.e., variables that were modified) as a function over the inputs (see \figref{sec_3.1_speclang}).
%\\[0.8em]
\paragraph{\textbf{Proof of Completeness.}} To understand that \sys's algorithm is complete with respect to the search space, we first show that that the algorithm always terminates. Recall that we use recursive bounds to finitize the number of solutions expressible by our IR's grammar. As explained in ~\secref{sec_4.1_verifierfailures}, we prevent the same solution from being regenerated, thus ensuring forward progress in search. These two facts imply that our algorithm always terminates. There are only two possible exit points for the {\tt while(true)} loop in our algorithm: line \ref{lst:nosols} and line~\ref{lst:solsfound} of \figref{sec_4.3_fullalgorithm}. The first is only reached once the entire search space has been exhausted. The second implies that a solution is successfully returned as $\Delta$ is not empty. It is important to note that our search algorithm is complete only for {\em verifiably correct} summaries. If a correct summary exists in the search space but cannot be proven correct using the available automated theorem prover, it will not be returned. Therefore, the completeness of the algorithm is modulo the completeness of the theorem prover.

\section{Intermediate Representation Specification}
\label{appendix:ir}
Here, we list the full set of types available in our IR and provide examples to demonstrate how they may be used to express models for library methods and types.

\begin{table}[ht]
\small
\centering
\begin{tabular}[t]{ |l|l| }
 \hline
 \multicolumn{2}{|c|}{\textbf{Primitive Data Types}}	\\
 \hline
 Scalars  				& {\tt bool}, {\tt int}, {\tt float}, {\tt string}, {\tt char}, ...	\\
\hline	
 Structures				& {\tt class(id:Type, id2:Type2, ..)} \\
\hline
 List					& {\tt list(Type)}             	\\
\hline
 Array				    & {\tt array(dimensions, Type)}		\\
\hline
 Functions			    & {\tt name(arg1:Type1, ...) : Type -> Body}	\\
\hline
 Conditionals		    & {\tt if $cond$ then $e_1$ else $e_2$}	\\
\hline
 Synthesis Construct    & {\tt choose($e_1$, $e_2$, ... , $e_n$)}	\\
\hline
\end{tabular}
\vspace{-0.4cm}
\end{table}

\begin{table}[ht]
\small
\centering
\begin{tabular}[t]{ |l|l| }
 \hline
 \multicolumn{2}{|c|}{\textbf{Built-in operations}}	\\
 \hline
 Arithmetic				& $+$, $-$, $*$, $/$, $\%$, ...	\\
\hline	
 Bitwise			    & $<<$, $>>$, $\&$, ... \\
\hline
 Relational				& $<$, $>$, $\leq$, $\geq$, ...              	\\
\hline
 Logical			    & $\&\&$, $\|$, $==$, $!=$		\\
\hline
 List			        & {\tt len, append, get, equals, concat, slice}	\\
\hline
 Array			        & {\tt select, store}	\\
\hline
\end{tabular}
\end{table}

To provide support for a datatype found in a Library, users must define the type of the object using our IR and annotate it with the fully qualified name, as follows:

\begin{lstlisting}[language=java,numbers=none]
@java.awt.Point
class Point(x:int, y:int)
\end{lstlisting}

Similarly, users may also provide support for library methods, for instance the following defines a model for the absolute value function:
\begin{lstlisting}[language=java,numbers=none]
@java.lang.Math.abs
abs(val: int) : int ->
  if val < 0 then val * -1 else val
\end{lstlisting}

Using the core IR described above, we implemented in \sys the $map$, $reduce$ and $join$ primitives used to synthesize summaries. We have also implemented commonly used methods from Java standard libraries such as {\tt java.util.Math,String,Date} and other essential data-types, along with methods that were needed to translate the Fiji plugins.

The $choose$ operator in the IR is a special construct that enables us to express a search space using the IR. The parameters to $choose$ are one or more expressions of matching types. The synthesizer is then free to select any expression from the list of choices in order to satisfy the correctness specification.

\if 0
\begin{lstlisting}[language=java,numbers=none,xleftmargin=0.1cm]
generator lambda_m(val:Element, depth:int) : list(Pair) ->
  if depth = 0 then 
    empty(Pair)
  else
    append(Pair(intExpr, intExpr), lambda_m(val, depth-1))
    
generator intExpr(val:Element, depth:int) : list(Pair) ->
  if depth = 0 then 
    empty(Pair)
  else
    append(Pair(intExpr, intExpr), lambda_m(val, depth-1))
\end{lstlisting}
\fi

\vfill\eject
\section{Code Generation Rules}
\label{appendix:codegen}
To generate target DSL code from the synthesized program summary, we implemented in \sys a set of translation rules that map the operators in our IR to the concrete syntax of the target DSL. Here, we list a subset of such code-generation rules for the Spark RDD API.

\begin{figure}[h]
\centering
\small
\begin{eqnarray*}
TR\tr{\textbf{map}(l, ~ \lambda_m : T \rightarrow list(Pair))} &=& \texttt{l.flatMapToPair(}\tr{\lambda_m} \texttt{);} \\
TR\tr{\textbf{map}(l, ~ \lambda_m : T \rightarrow list(U))} &=& \texttt{l.flatMap(}\tr{\lambda_m} \texttt{);} \\
TR\tr{\textbf{map}(l, ~ \lambda_m : T \rightarrow Pair)} &=& \texttt{l.mapToPair(}\tr{\lambda_m} \texttt{);} \\
TR\tr{\textbf{map}(l, ~ \lambda_m : T \rightarrow U)} &=& \texttt{l.map(}\tr{\lambda_m} \texttt{);} \\
TR\tr{\textbf{reduce}(l : list(Pair), ~ \lambda_r)}   &=& \texttt{l.reduceByKey(}\tr{\lambda_r} \texttt{);} \\
TR\tr{\textbf{reduce}(l : list(U), ~ \lambda_r)}   &=& \texttt{l.reduce(}\tr{\lambda_r} \texttt{);} \\
\hline
\vspace{0.05cm} \\
TR\tr{\lambda_m(e) \rightarrow e_b} &=& \texttt{(e -> } \tr{e_b} \texttt{)} \\
TR\tr{ e_1 + e_2 } &=& \tr{e_1} ~ \texttt{+} ~ \tr{e_2} \\
\end{eqnarray*}
\end{figure}

The translation function $TR$ takes as input an expression in our IR language and maps it to an equivalent expression in Spark. Since Spark provides multiple variations for the operators defined in our IR, such as $map$, we can select the appropriate variation by looking at the type information of the $\lambda_m$ function used by $map$. For example, if $\lambda_m$ returns a list of Pairs, we translate to {\tt JavaRDD.flatMapToPair}. If it instead returns a list of a non-Pair type, we use the more general rule that translates $map$ to \\
{\tt JavaRDD.flatMap}. Translation for the other expressions proceeds similarly.

\section{Program Analyzer Outputs}
\label{appendix:example}
Here, we use TPC-H Query 6 to illustrate the outputs computed by \sys's program analyzer. Since the queries are originally in SQL, we have manually translated them to Java as follows:
\begin{lstlisting}[language=java]
double query6(List<LineItem> lineitem){
  List<LineItem> lineitem = new ArrayList<LineItem>();
  Date dt1 = Util.df.parse("1993-01-01");
  Date dt2 = Util.df.parse("1994-01-01");
  double revenue = 0;
  for (LineItem l : lineitem) {
    if (
      l.l_shipdate.after(dt1) &&
      l.l_shipdate.before(dt2) &&
      l.l_discount >= 0.05 &&
      l.l_discount <= 0.07 &&
      l.l_quantity < 24
    )
      revenue += (l.l_extendedprice * l.l_discount);
  }
  return revenue;
}
\end{lstlisting}
First, \sys's program analyzer normalizes the loop starting on Line 6 into an equivalent {\tt while(true)\{..\}} loop, and then  traverses the loop to identify the set of input/output variables and operators used:

\begin{table}[ht]
\centering
\small
\begin{tabular}[t]{ |l|l| }
 \hline
 \multicolumn{2}{|c|}{\textbf{Program Analaysis Results}}	\\
 \hline
 Inputs Vars		& {\tt l: list(LineItem), dt1: Date, dt2: Date}	\\
\hline	
 Output Vars	    & {\tt revenue: double} \\
\hline
 Constants		        & {\tt [(24, int), (0.05, double), (0.07, double)]}	\\
\hline
 Operators				& $+$, $-$, $*$, $\geq$, $\leq$, $<$              	\\
\hline
 Methods			    & {\tt Date.before, Date.after}		\\
\hline
\end{tabular}
\end{table}

With this information, \sys generates verification conditions like those shown in ~\figref{rwm_vcs} for the row-wise mean benchmark. Next, the program analyzer defines a search space within which \sys searches for summaries and the needed loop-invariant. Since the full search space description is too large to show, we only show a small snippet below:

\begin{lstlisting}[language=java,numbers=none,xleftmargin=0.1cm]
generator doubleExpr(val:LineItem, depth:int) : double ->
  if depth = 0 then 
    choose(
      val.l_quantity, 
      val.l_extendedprice, 
      val.l_discount, 
      0.05, 
      0.07, 
      24
    )
  else
    choose(
      doubleExpr(val, 0), 
      doubleExpr(val, depth-1) + doubleExpr(val, depth-1), 
      doubleExpr(val, depth-1) * doubleExpr(val, depth-1), 
      doubleExpr(val, depth-1) / doubleExpr(val, depth-1)
    )
\end{lstlisting}

The {\tt doubleExpr} is the part of the grammar used to construct expressions that evaluate to {\em double}. The {\tt generator} keyword indicates that this is a special type of function, one that can select a different value from the {\tt choose} operators on each invocation. The depth parameter controls how large the generated expression is allowed to grow. The {\tt choose} construct is used to present a set of possible productions to the synthesizer. This grammar is tailored specifically to our implementation of TPC-H Query 6.

\section{Supplementary Experiments}
\label{appendix:experiments}

\subsection{Benchmark Details}
\label{appendix:features}
The benchmarks \sys extracted form a diverse and challenging problem set. As shown in the table below, they vary across programming style as well as the structure of their solutions. 
\begin{table}[h]
\centering
\begin{tabular}[h]{ |l|c|c| }
\hline
 \textbf{Benchmark Properties} &   \textbf{\# Extracted} & \textbf{\# Translated} 	\\
\hline
 Conditionals               	&    	26			   & 19	\\
\hline
 User Defined Types	 		    & 		14			   & 10	\\
\hline  
 Nested Loops                	&     	40			   & 22	\\
\hline
 Multiple Datasets              &     	22			   & 18	\\
\hline
 Multidim. Dataset      	    &    	38			   & 23	\\
\hline
\end{tabular}
\vspace{0.2cm}
%\caption{Properties of benchmarks translated by \sys.}
%\tablabel{variety}
\end{table}
\subsection{Developer Selection Criteria}
\label{appendix:developer}
To get reference Spark implementations for non-SQL benchmarks, we hired developers through the online freelancing platform UpWork.com. While hiring, we ensured all candidates met the following basic criteria:
\begin{compactenum}
\item At least an undergraduate or equivalent degree in computer science.
\item Minimum 500 hours of work logged at the platform.
\item Minimum 4 star rating for previous projects (scale of 5).
\item A portfolio of at least one or more successfully completed contracts using Spark.
\end{compactenum}
Finally, applicants were required to answer three test questions regarding Spark API internals to bid on our contract.

\subsection{Evaluating Cost Model Heuristics}
\label{appendix:costmodel}
We present here some experiments that measure whether \sys's cost model model can effectively identify efficient solutions during the search process.

\begin{table}[h]
\begin{tabular}{ |l|c|c|c| }
\hline
	\textbf{Program} & 
	\textbf{Emitted (MB)} & 	
	\textbf{Shuffled (MB)} 	&	
	\textbf{Runtime (s)} \\
\hline
WC 1 & 105k & 30 & 254 \\
\hline
WC 2 & 105k & 58k & 2627 \\
\hline
SM 1  & 16 & 0.7 & 189 \\
\hline
SM 2  & 90k & 0.7 & 362 \\
\hline
\end{tabular}

\if 0
\begin{tabular}{ |l|c|c|c|c| }
\hline
	\textbf{Program} & 
	\textbf{Data Emitted (MB)} & 	
	\textbf{Data Shuffled (MB)} 	&	
	\textbf{Runtime (s)} \\
\hline
\hline
WordCount (With Combiners)      & 105k & 30 & 254 \\
\hline
WordCount (Without Combiners)   & 105k & 58k & 2627 \\
\hline
\hline
StringMatch (Conditional Emit)  & 16 & 0.7 & 189 \\
\hline
StringMatch (Always Emit)    	& 90k & 0.7 & 362 \\
\hline
\end{tabular}
\fi
%\begin{table}
%    \begin{tabular}{ |l|c|c|c|c| }
%	\hline
%	\textbf{Version}	&	\textbf{Dataset} & \textbf{Data Emitted} & 
%	\textbf{Data Shuffled} 	&	\textbf{Runtime} \\
%	\hline
%	\multirow{3}{10em}{WordCount\\ (With Combiners)} 
%									& 25GB &  & 18.1MB & 1m41s \\
%	                                & 50GB &  & 25.7MB & 2m55s \\
%	 	                            & 75GB &  & 30.4MB & 4m14s \\
%	\hline
%	\multirow{3}{10em}{WordCount\\ (Without Combiners)}
%									& 25GB &  & 19.6GB & 12m42s \\
%	 		                        & 50GB &  & 41.3GB & 30m13s \\
%	 	                            & 75GB &  & 58.4GB & 43m47s \\
%	\hline
%	\end{tabular}
%   \caption{The correlation of data shuffle and execution time.}
%  \tablabel{costmodelproof1}
%\end{table}

\caption{The correlation of data shuffle and execution. (WC = WordCount, 
SM = StringMatch).}
\tablabel{costmodelproof1}
\vspace{-0.7cm}
\end{table}

As discussed in~\secref{sec_4.4_costmodel}, \sys uses a data-centric cost model. The cost model is based on the hypothesis that the amount of data generated and shuffled during the execution of a MapReduce program determines how fast the program executes. For our first experiment, we measured the correlation between the amount of data shuffled and the runtime of a benchmark to check the validity of the hypothesis. To do so, we compared the performance of two different Spark WordCount implementations: one that aggregates data locally before shuffling (WC 1) using combiners~\cite{mapreduce}, and one that does not (WC 2). Although both implementations processed the same amount of input data, the former implementation significantly outperformed the latter, as the latter incurred the expensive overhead of moving data across the network to the nodes responsible for processing it. \tabref{costmodelproof1} shows the amount of data shuffled along with the corresponding runtimes for both implementations using the 75GB dataset. As shown, the implementation that used combiners to reduce data shuffling was almost an order of magnitude faster.

Next, we verified the second part of our hypotheses by measuring the correlation of the amount of data generated and the runtime of a benchmark. To do so, we compared two solutions for the StringMatch benchmark (sequential code shown in~\figref{dynamic_orig}). The benchmark determines whether certain keywords exist in a large body of text. Both solutions use combiners to locally aggregate data before shuffling. However, one solution emits a key-value pair only when a matching word is found (SM 1), whereas the other always emits either {\tt (key, true)} or {\tt (key, false)} (SM 2). Since the data is locally aggregated, each node in the cluster only generates 2 records for shuffling (one for each keyword) regardless of how many records were emitted during the map phase. As shown in~\tabref{costmodelproof1}, the implementation that minimized the amount of data emitted in the map-phase executed almost twice as 
fast.

In sum, the two experiments confirm that the heuristics used in our cost model are accurate indicators of runtime performance for MapReduce applications. We also demonstrated the need for a data-centric cost model; solutions that minimize data costs execute significantly faster than those that do not.

\subsection{Evaluating Scalability of Generated Implementations}
\label{appendix:scalability}
\begin{figure}
    \input{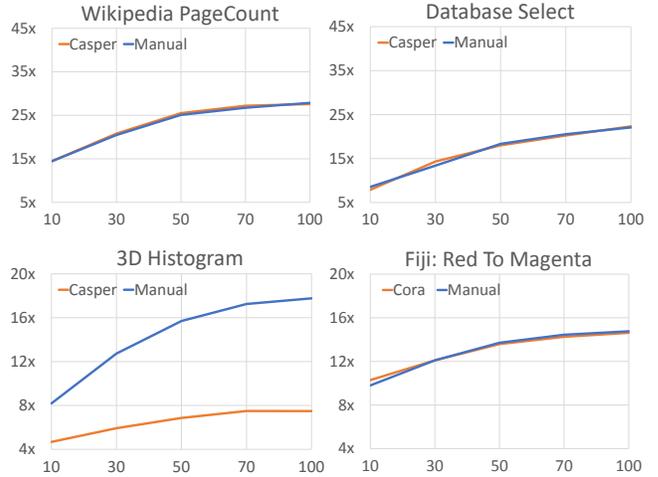}
    \vspace{-0.6cm}
    \caption{The top 2 benchmarks with the most performance along with the bottom 2. The x-axis plots the size of input data, while the y-axis plots the runtime speedup over sequential implementations.}
    \figlabel{speedups}
\end{figure}

To observe how implementations generated by \sys scale, we executed our benchmarks on different amounts of data and measured the resulting speedups. As shown in~\figref{speedups}, the \sys-generated Spark implementations exhibited good data parallelism and showed a steady increase in speedups across all translated benchmarks as the input data size increased, until the cluster reached maximum utilization.

\end{document}